\documentclass  [a4paper, 12pt] {article}
\usepackage{amsmath}
\usepackage{amssymb}
\usepackage{amsfonts}
\usepackage{amsthm}
\usepackage [latin1] {inputenc}
\usepackage [T1] {fontenc}
\usepackage[french]{babel}
\newtheorem {theo}{Theorem}[section]

\newtheorem {lem}{Lemma}[section]

\newtheorem {Rq}{Remark}[section]

\newcommand{\Tr}{\mathrm{Tr}}
\renewcommand{\Im}{\mathrm{Im}}

\makeatletter

\@addtoreset{equation}{section} \makeatother

\def\R{\mathbb{R}}
\def\C{\mathbb{C}}
\def\N{\mathbb{N}}
\def\Z{\mathbb{Z}}

\title{ ASYMPTOTIC PROPERTIES OF RANDOM MATRICES OF LONG-RANGE PERCOLATION
MODEL }
\author{{\bf S.Ayadi}\thanks{LMV - Laboratoire de Math\'ematiques
de Versailles, Universit\'e de Versailles Saint-Quentin-en-Yvelines,
78035 Versailles (FRANCE). E-mail: ayadi@math.uvsq.fr}}
\date{}
\begin{document}
\maketitle

{\bf Abstract}: We study the spectral properties of matrices of
long-range percolation model. These are $N\times N$ random real
symmetric matrices $H=\{H(i,j)\}_{i,j}$ whose elements are
independent random variables taking zero value with probability
$1-\psi\left( (i-j)/b\right)$, $b\in \mathbb{R}^{+}$, where $\psi$
is an even positive function with $\psi(t)\le{1}$ and vanishing at
infinity. We study the resolvent $G(z)=(H-z)^{-1}, \ Imz\neq{0}$ in
the limit $N,b\rightarrow\infty$, $b=O(N^{\alpha}), \ 1/3<\alpha<1$
and obtain the explicit expression $T(z_{1},z_{2})$ for the leading
term of the correlation function of the normalized trace of
resolvent $g_{N,b}(z)=N^{-1}Tr G(z)$. We show that in the scaling
limit of local correlations, this term leads to the expression
$(Nb)^{-1}T(\lambda+r_{1}/N+i0,\lambda+r_{2}/N-i0)=
b^{-1}\sqrt{N}|r_{1}-r_{2}|^{-3/2}(1+o(1))$ found earlier by other
authors for band random matrix ensembles. This shows that the ratio
$b^{2}/N$ is the correct scale for the eigenvalue density
correlation function and that the ensemble we study and that of band
random matrices belong to the same class of spectral universality.

\vskip0,5cm

 \vskip0,2cm

 \vskip0,2cm

{\bf running title}: {\it Asymptotic properties for percolation
model.}

\vskip0,5cm

\section{Introduction}
Random matrices play an important role in various fields of
mathemathics and physics. The eigenvalue distribution of large
matrices was initially considered by E.Wigner to model the
statistical properties of the energy spectrum of heavy nuclei (see
e.g. the collection of early papers \cite{BBB}). Further
investigations have led to numerous applications of random matrices
of infinite dimensions in such branches of theoretical physics as
statistical mechanics of disordered spin systems, solid state
physics, quantum chaos theory, quantum field theory and others (see
monographs and reviews \cite{I,II,III,IIII}). In mathematics, the
spectral theory of random matrices has revealed deep links with the
orthogonal polynomials theory, integrable systems, representation
theory, combinatorics, free probability theory, and others
\cite{BleIts,Deift,Soshni,Voi}.

 The first result of the spectral theory of large random matrices concerns the
eigenvalue distribution of the Wigner ensemble $A_{N}$ of $N\times
N$ real symmetric matrices of the form
\begin{equation}\label{b.1ANij}
A_{N}(i,j)=\frac{1}{\sqrt{N}}a(i,j), \quad |i|,|j|\le{n},
\end{equation}
where $N=2n+1$ and $\{a(i,j); \ -n\le{i}\le{j}\le{n}\}$ are
independent and identically distributed random variables defined on
the same probability space $(\Omega,\mathfrak{F},{\bf P})$ such that
\begin{equation}\label{b.1Eaij}
{\bf E}\{a(i,j)\}=0, \quad {\bf
E}\{a(i,j)^{2}\}=v^{2}(1+\delta_{ij}),
\end{equation}
where
\begin{equation*}
\delta_{ij}=\left\{
\begin{array}{lll}
0 & \textrm{if} & i\neq{j}, \\
1 &  \textrm{if} & i=j
\end{array}\right.
\end{equation*}
is the Kronecker symbol and ${\bf E}\{\cdot\}$ is the mathematical
expectation with respect to ${\bf P}$.

Denoting by $\lambda^{(n)}_{-n}\le{\ldots}\le{\lambda^{(n)}_{n}}$
the eigenvalues of $A_{N}$, the normalized eigenvalue counting
function is defined by
\begin{equation}\label{b.1sigma}
\sigma_{n}(\lambda,A_{N})=N^{-1}\sharp\{\lambda^{(n)}_{j}\le{\lambda}\}.
\end{equation}
E.Wigner \cite{BB} proved that if $a(i,j)$ has all order finite
moments, the eigenvalue counting measure
$d\sigma_{n}(\lambda,A_{N})$ converges weakly in average as
$n\rightarrow \infty$ to a distribution $d\sigma_{sc}(\lambda)$,
where the nondecreasing function $\sigma_{sc}(\lambda)$ is
differentiable and its derivative $\rho_{sc}$ is given by
\begin{equation}\label{b.1rho}
\rho_{sc}{(\lambda)}=\sigma_{sc}^{'}(\lambda)=\frac{1}{2\pi
v^{2}}\left\{
\begin{array}{lll}
\sqrt{4v^{2}-\lambda^{2}} & \textrm{if} &
|\lambda|\le{2\sqrt{v^{2}}}, \\
0                  &  \textrm{if}  &    |\lambda|>2v.
\end{array}\right.
\end{equation}
This limiting distribution (\ref{b.1rho}) is known as the Wigner
distribution, or the semicircle law. A proof of the Wigner's result
based on the resolvent technique is given in \cite{EEE,FFF,FF}.

 Important generalizations of the Wigner's ensemble are given by the
band and dilute random matrix ensembles \cite{CCC}. In the band
random matrices model, the matrix elements take zero value outside
the band of width $b_{n}$ along the principal diagonal, for some
positive sequence $(b_{n})_{n\geq{0}}$ of real numbers. This
ensemble can be obtained from $A_{N}$ (\ref{b.1ANij}) by multiplying
each $a(i,j)$ by $I_{(-1/2,1/2)}\left( (i-j)/b_{n}\right)$, where
\begin{equation*}
I_{B}(t) = \left\{
\begin{array}{lll}
1 & \textrm{if} & t\in{B}, \\
0 &  \textrm{if} & t\in\mathbb{R}\setminus B
\end{array}\right.
\end{equation*}
is the indicator function of the interval $B$. The ensemble of
dilute random matrices can be obtained from $A_{N}$ (\ref{b.1ANij})
by multiplying $a(i,j)$ by independent Bernoulli random variables of
parameter $ p_{n}/N$. Assuming that $b(n)=o(n)$ for large $n$, the
semicircle law is observed for both ensembles, in the limit
$b_{n}\rightarrow\infty$ (see \cite{AA}) and
$p_{n}\rightarrow\infty$ as $n\rightarrow\infty$ (see \cite{CCC}).

The crucial observation made numerically \cite{Casati} and then
supported in the theoretical physics (see \cite{Fyodorv,syl}) is
that the ratio $b^{2}/n$ is the critical one for the corresponding
transition in spectral properties of band random matrices. In
\cite{B}, it was proved that the ratio
$\tilde{\alpha}=\lim_{n\rightarrow\infty}b^{2}/n$ naturally arises
when one considers the leading term of this correlation function on
the local scale. This can be regarded as the support of the
conjecture that the local properties of spectra of band random
matrices depend on $\tilde{\alpha}$.

 Let us describe our results in more details. We are interested in a
generalization of the both ensembles mentioned above. Roughly
speaking, we consider the band random matrices with a random width.
To proceed, we consider the ensemble $\{H_{n,b}\}$ of random
$N\times N$ matrices, $N=2n+1$ whose entries $H_{n,b}$ is obtained
as follows: we multiply each matrix element $a(i,j)$ by some
Bernoulli random variable $d_{b}(i,j)$ with parameter $\psi\left(
(i-j)/b\right)$.

The family $\{d_{b}(i,j); \ |i|,|j|\le{n}\}$ can be regarded as the
adjacency matrix of the family of random graphs $\{\Gamma_{n}\}$
with $N=2n+1$ vertices $(i,j)$ such that the average number of edges
attached to one vertex is $b_{n}$. Hence, each edge $e(i,j)$ of the
graph is present with probability $\psi\left( (i-j)/b\right)$ and
not present with probability $1-\psi\left( (i-j)/b\right)$. Below
are some well known examples:
\begin{enumerate}
\item[$-$] In theoretical physics, the ensemble $\{\Gamma_{n}\}$ with $\psi(t)=e^{-|t|^{s}}$
is referred to as the Long-Range Percolation Model (see for example
\cite{K} and references therein). Our ensemble can be regarded as a
modification of the adjacency matrices of $\{\Gamma_{n}\}$. To our
best knowledge, the spectral properties of this model has not been
studied yet.
\item[$-$] It is easy to see that if one takes $b_{n}=N$ and $\psi\equiv 1$,
then one recovers the Wigner ensemble $(1.1)$.
\item[$-$]  If one considers $\psi(t)=I_{(-1/2,1/2)}(t)$,
one gets the band random matrix ensemble \cite{AA}.
\end{enumerate}
In present paper, we consider the resolvent
$G_{n,b}=\left(H_{n,b}-zI\right)^{-1}$ and study the asymptotic
expansion of the correlation function
\begin{equation*}
C_{n}(z_{1},z_{2})={\bf E}\{g_{n,b}(z_{1})g_{n,b}(z_{2})\}-{\bf
E}\{g_{n,b}(z_{1})\}{\bf E}\{g_{n,b}(z_{2})\},
\end{equation*}
where we denoted $g_{n,b}=N^{-1}\Tr G_{n,b}(z)$. Keeping $z_{l}$ far
from the real axis, we consider the leading term $T(z_{1},z_{2})$ of
this expansion and find explicit expression for it. This term
$T(r_{1}+i0,r_{2}-i0)$ regarded on the local scale $r_{1} -r_{2} =
r/N$ exhibits different behavior depending on the rate of decay of
the profile function $\psi(t)$.

Our main conclusion is that if $\psi(t)\sim |t|^{-1-\nu}$ as
$t\rightarrow\infty$, then the value $\nu=2$ separates two major
cases. If $\nu\in(1,2)$, then the limit of $T(r)$ depend on $\nu$.
If $\nu\in(2,+\infty)$, then
\begin{equation*}
\frac{1}{Nb}T(r)=-const \cdot \frac{\sqrt{N}}{b} \cdot
\frac{1}{|r|^{3/2}}(1+o(1)).
\end{equation*}
This asymptotic expression coincides with the result obtained in
\cite{B} for band random matrices. Then one can conclude that the
ensemble under consideration and the band random matrix ensemble
belong to the same universality class.

 The outline of this paper is as follows. In section 2, we define the
random matrix ensemble $H_{n,b}$ of long-range percolation model, we
state our main results and describe the scheme of their proofs. In
section 3, we study the correlation function $C_{n,b}(z_{1},z_{2})$
and obtain the main relation for it. In section 4, we show that
${\bf Var}\{g_{n,b}(z)\}$ is bounded by $(Nb)^{-1}$ and find the
leading term $T(z_{1},z_{2})$ of the correlation function under the
moment condition that $\sup_{i,j} {\bf E}|a(i,j)|^{14}<\infty$. In
section 5, we prove the auxiliary facts used in section 4.
Expressions derived in section 4 are analyzed in section 6, where
the asymptotic behavior of $T(z_{1},z_{2})$ is studied and the issue
of the universal bihaviour is discussed.

\section{The ensemble, main results and technical tools}
\subsection{The ensemble and main results}
Let us consider a family of independent real random variables ${\cal
A}_{n}= \{a(i,j); \ |i|, |j|\le{n}\}$ satisfying (\ref{b.1ANij}).
Let $\psi(t)$, $t\in\mathbf{R}$, be a real continuous even function
such that:
\begin{equation}\label{b.2psicondit}
0\le{\psi{(t)}}\le{1}, \quad  \int_{\mathbb{R}}\psi{(t)}dt = 1.
\end{equation}
Given real $b>0$, we introduce a family of independent Bernoulli
random variables ${\cal D}_{b}=\{d_{b}(i,j); \ |i|, |j|\le{n}\}$
with the law
\begin{equation}\label{b.2dijb}
d_{b}(i,j) = \left\{
\begin{array}{lll}
1 & \textrm{with probability} & \psi\left( (i-j)/b\right) \\
0 &  \textrm{with probability} & 1-\psi\left( (i-j)/b\right).
\end{array}\right.
\end{equation}
This family is independent of the family ${\cal A}_{n}$. We assume
that ${\cal A}_{n}$ and ${\cal D}_{n}$ are defined on the same
probability space $(\Omega,\mathfrak{F},{\bf P})$ and we denote by
${\bf E}\{\cdot\}$ the mathematical expectation with respect to
${\bf P}$.

We define a real symmetric $N\times N$ random matrix $H_{n,b}$ by
equality:
\begin{equation}\label{b.2Hnbij}
H_{n,b}(i,j)=\frac{1}{\sqrt{b}}a(i,j)d_{b}(i,j), \quad i\le{j},
\quad |i|, |j|\le{n},
\end{equation}
where $b\le{N}$, $N=2n+1$. Here and below the family $\{H_{n,b}\}$
is referred to as the ensemble of random matrices of long-range
percolation model. In what follows, we will need the existence of
several absolute moments of $a(i,j)$ that we denote by
\begin{equation}\label{b.2mur}
\mu_{l}=\sup_{|i|, |j|\le{n}}{\bf E}\{|a(i,j)|^{l}\},
\end{equation}
where the upper bound for $l$ will be specified later.

We consider the resolvent
\begin{equation*}
G_{n,b}(z)=(H_{n,b}-z)^{-1}, \quad \Im z\neq{0}.
\end{equation*}
Its normalized trace $g_{n,b}(z)$ coincides with the Stieltjes
transform of the normalized eigenvalue counting function
$\sigma_{n,b}(\lambda;H_{n,b})$ (\ref{b.1sigma}):
\begin{equation}\label{b.2gnbz}
g_{n,b}(z)=\frac{1}{N}\Tr G_{n,b}(z)=\int
(\lambda-z)^{-1}d\sigma_{n,b}(\lambda,H_{n,b}),
 \ \Im z \neq{0}.
\end{equation}
In \cite{A}, we have proved that if $\mu_{3}< \infty$ (\ref{b.2mur})
and $1\ll b\ll N$, then
\begin{equation*}
\lim_{n,b\rightarrow\infty}{\bf E}\{g_{n,b}(z)\}=w(z)
\end{equation*}
for $z\in \Lambda_{\eta}$, where
\begin{equation}\label{b.2Lambdaeta}
\Lambda_{\eta}=\{z\in\mathbf{ C}: \  \eta\le{|\Im z|}\}, \quad
\eta=2v+1
\end{equation}
and the limiting function $w(z)$ verifies equation
\begin{equation}\label{b.2wz}
w(z)=\frac{1}{-z-v^{2} w(z)}
\end{equation}
with $v$ is determined by (\ref{b.1Eaij}). Equation (\ref{b.2wz})
has a unique solution in the class of functions such that $\Im
w(z)\Im z>0$, $\Im z\neq{0}$. This solution $w(z)$ is the Stieltjes
transform of the semi-circle distribution (\ref{b.1rho}). This
result shows that the semi-circle law is valid for random matrices
of long-range percolation model. As a by-product of proof, we have
shown that
\begin{equation}\label{b.2Vargnbz}
{\bf Var}\{g_{n,b}(z)\}=o(1), \ \quad z\in\Lambda_{\eta}, \quad
\hbox{ as } \quad n,b\rightarrow\infty
\end{equation}
and that the convergence $g_{n,b}(z)\rightarrow{w(z)}$ holds in
probability.

In this paper, we improve the result (\ref{b.2Vargnbz}) in two
stages. On the first one we show that ${\bf
Var}\{g_{n,b}(z)\}=O\left((Nb)^{-1}\right)$ in the limit
$n,b\rightarrow\infty$ such that
\begin{equation}\label{b.2bnalpha}
b=O\left(n^{\alpha}\right),  \quad 1/3<\alpha<1
\end{equation}
and this gives the convergence $g_{n,b}(z)\rightarrow{w(z)}$ with
probability 1. Next, we find the explicit form of the leading term
of the correlation function
\begin{equation*}
C_{n,b}(z_{1},z_{2})={\bf E}\{g_{n,b}(z_{1})g_{n,b}(z_{2})\}-{\bf
E}\{g_{n,b}(z_{1})\}{\bf E}\{g_{n,b}(z_{2})\}.
\end{equation*}
We now formulate the main result of the paper, where we denote
$w_{1}=w(z_{1})$ and $w_{2}=w(z_{2})$ are given by (\ref{b.2wz}).

\vskip0,2cm

\begin{theo} Let ${\cal A}_{n}$ be such that, in addition
to (\ref{b.1Eaij}), the following properties are verified :
\begin{equation}\label{b.2Eaij3}
{\bf E}\{a(i,j)^{3}\}={\bf E}\{a(i,j)^{5}\}=0, \quad {\bf
E}\{a(i,j)^{2m}\}=V_{2m}(1+\delta_{ij})^{m}, \quad m=2,3
\end{equation}
for all $i\le{j}$, $\mu_{14}<\infty$ (\ref{b.2mur}) and
\begin{equation*}
\int_{\R} \sqrt{\psi(t)}dt<\infty.
\end{equation*}
Then in the limit $n,b\rightarrow\infty$ (\ref{b.2bnalpha}) and for
$z_{l}\in\Lambda_{\eta}$ (\ref{b.2Lambdaeta}), $l=1,2$, equality
\begin{equation}\label{b.2Cnbz1z2Th}
C_{n,b}(z_{1},z_{2})
=\frac{1}{Nb}T(z_{1},z_{2})+o\left(\frac{1}{Nb}\right)
\end{equation}
holds with $T$ is given by the formula
\begin{equation}\label{b.2Tz1z2}
T(z_{1},z_{2})=Q(z_{1},z_{2})+\frac{2\Delta w_{1}^{3}w_{2}^{3}}{(1 -
v^{2}w_{1}^{2})(1-v^{2}w_{2}^{2})}
\end{equation}
with
\begin{equation}\label{b.2Qz1z2}
Q(z_{1},z_{2})=\frac{v^{2}w_{2}^{2}w_{1}^{2}}{\pi(1-v^{2}w_{1}^{2})(
1- v^{2}w_{2}^{2})} \int_{\R}\frac{\tilde{\psi}(p)}{[ 1-
v^{2}w_{1}w_{2}\tilde{\psi}(p)]^{2}}dp,
\end{equation}
where $\tilde{\psi}(p)$ is the Fourier transform of $\psi$
\begin{equation*}
\tilde{\psi}(p)=\int_{\R}\psi(t)e^{ipt}dt
\end{equation*}
and
\begin{equation}\label{b.2Delta}
\Delta= V_{4}\int_{\R}\psi(t)dt-3v^{4}\int_{\R}\psi^{2}(t)dt.
\end{equation}
\end{theo}

\vskip0,5cm

Under this conditions, Theorem 2.1 and relation (\ref{b.2Tz1z2})
remain true with $\Delta$ replaced by
\begin{equation*}
\lim_{n,b\rightarrow \infty} \sup_{|i|\le{n}}\left(
b\sum_{|j|\le{n}} {\bf E}\{H(i,j)^{4}\} - 3 {\bf
E}\{H(i,j)^{2}\}^{2}\right)
\end{equation*}
\begin{equation*}
=\lim_{n,b\rightarrow \infty}
\sup_{|i|\le{n}}\left(\frac{1}{b}\sum_{|j|\le{n}}
(1+\delta_{ij})^{2}\left[V_{4}\psi(\frac{i-j}{b})
-3v^{4}\psi(\frac{i-j}{b})^{2}\right]\right).
\end{equation*}
We would like to note that the form of (\ref{b.2Tz1z2}) generalizes
the expressions obtained in \cite{B} and \cite{C}. Namely, the term
$Q(z_{1},z_{2})$ is derived for the case when the entries of random
matrices $H$ are gaussian random variables. The ensemble we consider
is very similar to the band random matrices, but it represents a
different model. The form of the last term is exactly the same as
the one obtained in \cite{C} for the Wigner random matrices. This
shows that this term " forgets " the band-like structure of our
matrices. All our computations and formulas are valid in the case of
band random matrices
$H_{n,b}(i,j)=b^{-1/2}a(i,j)[\psi\left((i-j)/b\right)]^{1/2}$ with
not necessarily gaussian $a(i,j)$. Therefore Theorem 2.1 generalizes
the results of paper \cite{B}. In the case of band random matrices,
one obtains the same expressions (\ref{b.2Cnbz1z2Th}) and
(\ref{b.2Tz1z2}) with $\Delta$ (\ref{b.2Delta}) replaced by
$\Delta_{band}=(V_{4}-3v^{4})\int \psi^{2}(t)dt$, provided $a(i,j)$
are the same as in Theorem 2.1.

The results of Theorem 2.1 are used to study the universality
properties of eigenvalue distribution. We do this in Section 6.

\vskip0,5cm
\subsection{Cumulant expansions and resolvent identities}
We prove Theorem 2.1 and Theorem 2.2 by using the method proposed in
papers \cite{C,CCC} and further developed in a series of works
\cite{A,B}. The basic tools of this method are given by the
resolvent identities combined with the cumulant expansions
technique.
\subsubsection{The cumulant expansions formula} Let us consider a
family $\{X_{t}: \ t=1,\ldots,m\}$ of independent real random
variables defined on the same probability space such that ${\bf
E}\{|X_{t}|^{q+2}\}<\infty $ for some $q\in \N$ and $t=1,\ldots,m$.
Then for any complex-valued function $F(u_{1},\ldots,u_{m})$ of the
class $\mathcal{C}_{\infty}(\R^{m})$ and for all $j$, one has
\begin{equation}\label{b.2EXjFX1Xm}
{\bf E}\{X_{t} F(X_{1},\ldots,X_{m})\}=\sum_{r=0}^{q}
\frac{K_{r+1}}{r!} {\bf
E}\left\{\frac{\partial^{r}F(X_{1},\ldots,X_{m})}
{(\partial{X_{t}})^{r}}\right\} +\epsilon_{q}(X_{t}),
\end{equation}
where $K_{r}=Cum_{r}(X_{t})$ is the r-th cumulant of $X_{t}$ and the
remainder $\epsilon_{q}(X_{t})$ can be estimated by inequality
\begin{equation}\label{b.2epsilonqXj}
|\epsilon_{q}(X_{t})|\le{C_{q} \sup_{U\in \R^{m}}
\left|\frac{\partial^{q+1}F(U)}{\partial{u_{t}^{q+1}}}\right|{\bf
E}\{|X_{t}|^{q+2}\}},
\end{equation}
where $C_{q}$ is a constant. Relations (\ref{b.2EXjFX1Xm}) and
(\ref{b.2epsilonqXj}) can be proved by multiple using of the
Taylor's formula (see \cite{A,C} for the proofs).

\begin{Rq} The cumulants $K_{r}$ can be expressed in terms of the moments
$\breve{\mu}_{r}={\bf E}(X_{t}^{r})$ of $X_{t}$.

Indeed, let $f_{t}$ be a complex-valued function of one real
variable such that
\begin{equation*}
f_{t}(x)=F(X_{1},\ldots,X_{t-1},x,X_{t+1},\ldots,X_{n})
\end{equation*} and $f_{t}^{(r)}$ is its r-th derivative.

 \vskip0,2cm
\begin{itemize}
\item[$\bullet$] If $q=1$ and ${\bf E}\{X_{t}\}=0$, then
\begin{equation}\label{c.K1K2}
K_{1}=\breve{\mu}_{1}=0, \quad K_{2}=\breve{\mu}_{2}
\end{equation}
and the remainder $\epsilon_{1}(X_{t})$ is given by:
\begin{equation}\label{b.2epsilon1Xj}
\epsilon_{1}(X_{t})=\frac{1}{2}{\bf
E}\left\{X_{t}^{3}f_{t}^{(2)}(x_{0})\right\}-K_{2}{\bf
E}\left\{X_{t}f_{t}^{(2)}(x_{1})\right\}.
\end{equation}
\item[$\bullet$] If $q=3$ and ${\bf E}\{X_{t}\}= {\bf
E}\{X_{t}^{3}\}=0$, then
\begin{equation}\label{b.2K2K4}
K_{1}=K_{3}=0, \quad K_{2}=\breve{\mu}_{2}, \quad
K_{4}=\breve{\mu}_{4}-3\breve{\mu}^{2}_{2}
\end{equation}
and the remainder $\epsilon_{3}(X_{t})$ is given by:
\begin{align}
\nonumber \epsilon_{3}(X_{t})=& \frac{1}{4!}{\bf
E}\left\{X_{t}^{5}f_{t}^{(4)}(x_{0})\right\}-\frac{K_{2}}{3!}{\bf
E}\left\{X_{t}^{3}f_{t}^{(4)}(x_{1})\right\}\\
\label{b.2epsilon3Xj} &-\frac{K_{4}}{3!}{\bf
E}\left\{X_{t}f_{t}^{(4)}(x_{2})\right\}.
\end{align}
\item[$\bullet$] If $q=5$ and ${\bf E}\{X_{t}\}= {\bf E}\{X_{t}^{3}\}={\bf
E}\{X_{t}^{5}\}=0$, then the cumulants $K_{r}$, $r=1,\ldots,4$ are
given by (\ref{b.2K2K4}),
\begin{equation}\label{b.2K5K6}
K_{5}=0, \quad
K_{6}=\breve{\mu}_{6}-15\breve{\mu}_{4}\breve{\mu}_{2}
+30\breve{\mu}^{3}_{2}
\end{equation}
and the remainder $\epsilon_{5}(X_{t})$ is given by:
\begin{align}
\nonumber \epsilon_{5}(X_{t})= &\frac{1}{6!}{\bf
E}\left\{X_{t}^{7}f_{t}^{(6)}(x_{0})\right\}-\frac{K_{2}}{5!}{\bf
E}\left\{X_{t}^{5}f_{t}^{(6)}(x_{1})\right\}\\
&\label{b.2epsilon5Xj} -\frac{K_{4}}{(3!)^{2}}{\bf
E}\left\{X_{t}^{3}f_{t}^{(6)}(x_{2})\right\}-\frac{K_{6}}{5!}{\bf
E}\left\{X_{t}f_{t}^{(6)}(x_{3})\right\},
\end{align}
where for each $\nu=0,\ldots,3$, $x_{\nu}$ is a real random variable
that depends on $X_{t}$ and such that $|x_{\nu}|\le{|X_{t}|}$. In
what follows, we denote $f_{t}^{(r)}(x_{\nu})
=[\partial^{r}{F}/\partial{X_{t}^{r}}]^{(\nu)}$.
\end{itemize}
\end{Rq}
\vskip0,1cm

\subsubsection{Resolvent identities} For any two real symmetric
$n \times n$ matrices $h$ and $\tilde{h}$ and any non-real $z$ the
resolvent identity
\begin{equation}\label{b.2h-zI-1}
( h-zI)^{-1}=(\tilde h-zI)^{-1}- ( h-zI)^{-1}( h-\tilde h)(\tilde
h-zI)^{-1}
\end{equation}
is valid. Regarding (\ref{b.2h-zI-1}) with, $\tilde h=0$ and
denoting $G=\left(h-zI\right)^{-1}$, we get equality
\begin{equation}\label{b.2Gij}
G(i,j) = \zeta\delta_{ij} - \zeta\sum_{s=1}^{n} G(i,s)h(s,j), \quad
\zeta={-z^{-1}},
\end{equation}
where $h(i,j), \ i,j=1,\ldots,n$ are the entries of the matrix $h$,
$G(i,j)$ are the entries of the resolvent $G$ and $\delta$ denotes
the Kronecker symbol.

Using (\ref{b.2h-zI-1}) we derive for $G=\left(h-zI\right)^{-1}$,
$|\Im z|\neq{0}$ equality
\begin{equation}\label{b.2partialG}
\frac{\partial{G(s,t)}}{\partial{h(j,k)}}=
-\frac{1}{1+\delta_{jk}}\left[G(s,j)G(k,t)+ G(s,k)G(j,t)\right].
\end{equation}
We will also need two more formulas based on (\ref{b.2partialG});
these are expressions for
$\partial^{2}{G(i,j)}/\partial{h(j,i)^{2}}$ and
$\partial^{3}{G(i,j)}/\partial^{3}{h(j,i)}$. We present them later.
\vskip0,1cm
\subsubsection{The scheme of the proof of Theorem 2.1}
In this subsection we present a schema of computation of the leading
terms of $C_{n,b}(z_{1},z_{2})$ (cf. (\ref{b.2Cnbz1z2Th})).

Let us denote $g_{l}=g_{n,b}(z_{l})$, $l=1,2$ (everywhere below, we
omit the subscripts $n$,$b$ when no confusion can arise). For a
given a random variable, we denote $\xi^{0}=\xi-{\bf E}\xi$. Then
using identity
\begin{equation}\label{b.2Ef0g0}
{\bf E}\{\xi^{0}g^{0}\}={\bf E}\{\xi^{0}g\},
\end{equation}
we rewrite $C_{12} =C_{n,b}(z_{1},z_{2})$ as
\begin{equation*}
C_{12}={\bf E}\{g^{0}_{1}g_{2}\}=
\frac{1}{N}\sum_{|i|\le{n}}R_{12}(i)
\end{equation*}
with $R_{12}(i)={\bf E}\{g^{0}_{1}G_{2}(i,i)\}$. Applying the
resolvent identity (\ref{b.2h-zI-1}) to $G_{2}(i,i)$, we obtain
equality
\begin{equation}\label{b.2R12}
R_{12}(i)=-\zeta_{2}\sum_{|p|\le{n}}{\bf
E}\{g^{0}_{1}G_{2}(i,p)H(p,i)\}.
\end{equation}
To compute ${\bf E}\{g^{0}_{1}G_{2}(i,p)H(p,i)\}$, we use the
cumulants expansion method (\ref{b.2EXjFX1Xm}), and get
\begin{align}
\nonumber {\bf E}\{g^{0}_{1}G_{2}(i,p)H(p,i)\}=& K_{2}{\bf
E}\left\{\frac{\partial\left(g^{0}_{1}G_{2}(i,p)\right)}
{\partial{H(p,i)}}\right\}\\
\label{b.2Eg0G2ipHpi} &+\frac{K_{4}}{6}{\bf
E}\left\{\frac{\partial^{3}\left(g^{0}_{1}G_{2}(i,p)\right)}
{\partial{H(p,i)^{3}}}\right\}+\tau_{ip},
\end{align}
where $K_{r}$ is the r-th cumulant of $H(p,i)$ and $\tau_{ip}$
vanishes. Substituting this equality in (\ref{b.2R12}) and using
(\ref{b.2partialG}), we obtain that
\begin{align}
\nonumber \frac{\partial\{g^{0}_{1}G_{2}(i,p)\}}{\partial{H(p,i)}}=&
g^{0}_{1}\frac{\partial G_{2}(i,p)}
{\partial{H(p,i)}}+G_{2}(i,p)\frac{1}{N}\sum_{|s|\le{n}}\frac{\partial
G_{1}(s,s)}{\partial{H(p,i)}}\\
\nonumber =&
-\frac{1}{1+\delta_{pi}}g^{0}_{1}[G_{2}(i,p)^{2}+G_{2}(i,i)G_{2}(p,p)]\\
\label{b.2partialg10G2ij} &
-\frac{1}{1+\delta_{pi}}\left\{\frac{2}{N}G^{2}_{1}(i,p)G_{2}(i,p)\right\},
\end{align}
where we used (\ref{b.2partialG}) in the form
\begin{equation*}
\frac{\partial\{g^{0}_{1}G_{2}(i,p)\}}{\partial{H(p,i)}}=
\left\{\frac{\partial\left(g^{0}_{1}G_{2}(i,p)\right)}
{\partial{h(p,i)}}\vert_{h=H}\right\}.
\end{equation*}
We get relation
\begin{align}
\nonumber R_{12}(i)= & \zeta_{2}v^{2}{\bf
E}\left\{g^{0}_{1}G_{2}(i,i)\sum_{|p|\le{n}}G_{2}(p,p)
\frac{1}{b}\psi\left(\frac{p-i}{b}\right)\right\}\\
\nonumber &+\frac{\zeta_{2}v^{2}}{b}\sum_{|p|\le{n}}{\bf
E}\{g^{0}_{1}G_{2}(i,p)^{2}\}\psi\left(\frac{p-i}{b}\right)\\
\nonumber &+ \frac{2\zeta_{2}v^{2}}{Nb}\sum_{|p|\le{n}}{\bf
E}\{G_{1}^{2}(i,p)G_{2}(i,p)\}\psi\left(\frac{p-i}{b}\right)\\
\label{b.2R12form1} &-\frac{\zeta_{2}}{6}\sum_{|p|\le{n}}K_{4}{\bf
E}\left\{\frac{\partial^{3}(g^{0}_{1}G_{2}(i,p)
)}{\partial{H(p,i)^{3}}}\right\}+\Phi_{n,b}(i),
\end{align}
where $\sup_{|i|\leq{n}}|\Phi_{n,b}(i)|$ vanishes as
$n,b\rightarrow\infty$ (\ref{b.2bnalpha}) (see subsection 3.2 for
more details). Also we have taken into account that (cf.
(\ref{b.2K2K4}))
\begin{equation*}
K_{2}(p,i)=K_{2}\left(H_{n,b}(p,i)\right)=\frac{1}{b}{\bf
E}\{a(p,i)^{2}d_{n,b}(p,i)^{2}\}=\frac{v^{2}}{b}
\psi\left(\frac{p-i}{b}\right)(1+\delta_{pi}).
\end{equation*}
Let us return to relation (\ref{b.2R12form1}). We observe that the
first term of the right-hand side (RHS) can be expressed in terms of
$R_{12}$. This gives the possibility to obtain an equation of
$R_{12}$. The second term vanishes in the limit
$n,b\rightarrow\infty$ (we give later the explicit formulation). The
third term represents the leading term of the correlations function
(which provides the first expression of (\ref{b.2Tz1z2})). The
fourth term gives the contribution of the order $O((Nb)^{-1})$ to
(\ref{b.2Cnbz1z2Th}) (which provides the second expression of the
leading term (\ref{b.2Tz1z2})). The last term $\Phi_{n,b}(i)$ gives
the contribution of the order $o((Nb)^{-1})$ to (\ref{b.2Cnbz1z2Th})
(see Lemma 3.2).

\section{Correlation function of the resolvent}
In this section we give the main relation of the correlation
function $C_{n,b}(z_{1},z_{2})$. In what follows, we will need two
elementary inequalities
\begin{equation}\label{b.3GijleImz}
|G(i,p)|\le{||G||}\le{\frac{1}{|\Im z|}},
\end{equation}
and
\begin{equation}\label{b.3sumGij2leImz2}
\sum_{|p|\le{n}}|G(i,p)|^{2}=||G\vec{e}_{i}||^{2}\le{\frac{1}{|\Im
z|^{2}}}, \quad |i|\le{n}
\end{equation}
that hold for the resolvent of any real symmetric matrix. Here and
below we consider $||e||_{2}^{2}=\sum_{i}|e(i)|^{2}$  and denote
by $||G||=\sup_{||e||_{2}=1}||Ge||_{2}$ the corresponding operator norm.\\

\subsection{Derivation of relations for $R_{12}(i)$} Let us consider
the average ${\bf E}\{g^{0}_{1}G_{2}(i,p)H(p,i)\}$. For each pair
$(i,p)$, $g^{0}_{1}G_{2}(i,p)$ is a smooth function of $H(p,i)$. Its
derivatives are bounded because of equation (\ref{b.2partialG}) and
(\ref{b.3GijleImz}). In particular
\begin{equation*}
|D^{6}_{pi}\{\hat{g}^{0}_{1}\hat{G}_{2}(i,p)\}| \leq{C\left(|\Im
z_{1}|^{-1} + |\Im z_{2}|^{-1}\right)^{8} },
\end{equation*}
where $C$ is an absolute constant. Here and thereafter we use the
notation $D_{pi}$ for $\partial/\partial{H(p,i)}$.

  According to the definition of $H$ and the condition
$\mu_{7}<\infty$ (\ref{b.2mur}), the seven absolute moment of
$H(p,i)$ is of order $1/(b^{7/2})$. Then we can apply
(\ref{b.2EXjFX1Xm}) with $q=5$ to ${\bf
E}\{g^{0}_{1}G_{2}(i,p)H(p,i)\}$ and using (\ref{b.2h-zI-1}), we get
relation
\begin{itemize}
\item[$\bullet$] if $p<i$
\begin{align}
\nonumber {\bf E}\{g^{0}_{1}G_{2}(i,p)H(p,i)\}= &
K_{2}\left(H(p,i)\right){\bf
E}\left\{D^{1}_{pi}\left(g^{0}_{1}G_{2}(i,p)\right)\right\} \\
\nonumber &+\frac{K_{4}\left(H(p,i)\right)}{6}{\bf
E}\left\{D^{3}_{pi}\left(g^{0}_{1}G_{2}(i,p)\right)\right\}\\
\label{b.3pi} &+\frac{K_{6}\left(H(p,i)\right)}{120}{\bf
E}\left\{D^{5}_{pi}\left(g^{0}_{1}G_{2}(i,p)\right)\right\}+
\tilde{\epsilon}_{pi}
\end{align}
with
\begin{align}
\nonumber \tilde{\epsilon}_{pi}= &\frac{1}{6!}{\bf
E}\left\{H(p,i)^{7}[D^{6}_{pi}(g^{0}_{1}G_{2}(i,p))]^{(0)}\right\}\\
\nonumber &-\frac{K_{2}\left(H(p,i)\right)}{5!}{\bf
E}\left\{H(p,i)^{5}[D^{6}_{pi}(g^{0}_{1}G_{2}(i,p))]^{(1)}\right\}\\
\nonumber &-\frac{K_{4}\left(H(p,i)\right)}{(3!)^{2}}{\bf
E}\left\{H(p,i)^{3}[D^{6}_{pi}(g^{0}_{1}G_{2}(i,p))]^{(2)}\right\}\\
\label{b.3tildepsilonpi} &-\frac{K_{6}\left(H(p,i)\right)}{5!}{\bf
E}\left\{H(p,i)[D^{6}_{pi}(g^{0}_{1}G_{2}(i,p))]^{(3)}\right\},
\end{align}
where the cumulants are given by (cf.
(\ref{b.2K2K4})-(\ref{b.2K5K6}))
\begin{equation}\label{b.3K2K4}
K_{2}\left(H(p,i)\right)=
\frac{v^{2}}{b}\psi(\frac{p-i}{b})(1+\delta_{pi}), \quad
K_{4}\left(H(p,i)\right)=\frac{\Delta_{pi}}{b^{2}}(1+\delta_{pi})^{2}
\end{equation}
with $\Delta_{pi}=
V_{4}\psi\left((p-i)/b\right)-3v^{4}\psi\left((p-i)/b\right)^{2}$
and
\begin{equation}\label{b.3K6}
K_{6}\left(H(p,i)\right)=\frac{\theta_{pi}}{b^{3}}(1+\delta_{pi})^{3},
\end{equation}
with $\theta_{pi} =
V_{6}\psi\left((p-i)/b\right)-15V_{4}v^{2}\psi\left((p-i)/b\right)^{2}
+30v^{6}\psi\left((p-i)/b\right)^{3}$. In (\ref{b.3tildepsilonpi}),
we have denoted for each pair $(p,i)$
\begin{equation*}
[g^{0}_{1}G_{2}(i,p)]^{(\nu)}=\{g^{(\nu)}\}^{0}_{pi}(z_{1})
G^{(\nu)}_{pi}(i,p;z_{2}), \quad \nu=0,\ldots,3
\end{equation*}
and $G^{(\nu)}_{pi}(z_{l})=(H^{(\nu)}_{pi}-z_{l})^{-1}$, $l=1,2$
with real symmetric
\begin{equation*}
H^{(\nu)}_{pi}(r,s)= \left\{
\begin{array}{lll}
H(r,s) & \textrm{if} & (r,s)\neq(p,i); \\
H^{(\nu)}(p,i) &  \textrm{if} & (r,s)=(p,i),
\end{array}\right.
\end{equation*}
where $|H^{(\nu)}(p,i)|\le{|H(p,i)|}$, $\nu=0,\ldots,3$ (see
subsection 2.2.1 for more detail).
\item[$\bullet$] If $i<p$, then using equality $H(p,i)=H(i,p)$, we
get
\begin{align}
\nonumber {\bf E}\{g^{0}_{1}G_{2}(i,p)H(i,p)\}=
&K_{2}\left(H(i,p)\right){\bf
E}\left\{D^{1}_{ip}\left(g^{0}_{1}G_{2}(i,p)\right)\right\}\\
\nonumber & + \frac{K_{4}\left(H(i,p)\right)}{6}{\bf
E}\left\{D^{3}_{ip}\left(g^{0}_{1}G_{2}(i,p)\right)\right\}\\
\label{b.3ip} &+\frac{K_{6}\left(H(i,p)\right)}{120}{\bf
E}\left\{D^{5}_{ip}\left(g^{0}_{1}G_{2}(i,p)\right)\right\}+
\tilde{\tilde{\epsilon}}_{ip},
\end{align}
where $\tilde{\tilde{\epsilon}}_{ip}$ is given by
(\ref{b.3tildepsilonpi}) with replaced $D_{pi}$ by $D_{ip}$ and
$K_{r}$ are the cumulants of $H(i,p)$ as in
(\ref{b.3K2K4})-(\ref{b.3K6}).
\item[$\bullet$] If $p=i$, then
\begin{align}
\nonumber {\bf E}\{g^{0}_{1}G_{2}(i,i)H(i,i)\}=
&K_{2}\left(H(i,i)\right){\bf
E}\left\{D^{1}_{ii}\left(g^{0}_{1}G_{2}(i,i)\right)\right\}\\
\nonumber &+ \frac{K_{4}\left(H(i,i)\right)}{6}{\bf
E}\left\{D^{3}_{ii}\left(g^{0}_{1}G_{2}(i,i)\right)\right\}\\
\label{b.3ii} &+\frac{K_{6}\left(H(i,i)\right)}{120}{\bf
E}\left\{D^{5}_{ii}\left(g^{0}_{1}G_{2}(i,i)\right)\right\}+
\tilde{\tilde{\tilde{\epsilon}}}_{ii},
\end{align}
where $\tilde{\tilde{\tilde{\epsilon}}}_{ii}$ is given by
(\ref{b.3tildepsilonpi}) with replaced $D_{pi}$ by $D_{ii}$ and
$K_{r}$ are the cumulants of $H(i,i)$ as in
(\ref{b.3K2K4})-(\ref{b.3K6}).
\end{itemize}
Substituting (\ref{b.3pi}), (\ref{b.3ip}) and (\ref{b.3ii}) into
(\ref{b.2R12}) and using (\ref{b.2partialg10G2ij}), we obtain
equality
\begin{align}
\nonumber R_{12}(i)= &\zeta_{2}v^{2}{\bf
E}\left\{g^{0}_{1}G_{2}(i,i)\sum_{|p|\le{n}}G_{2}(p,p)
\frac{1}{b}\psi\left(\frac{p-i}{b}\right)\right\}\\
\nonumber &+\frac{\zeta_{2}v^{2}}{b}\sum_{|p|\le{n}}{\bf
E}\{g^{0}_{1}G_{2}(i,p)^{2}\}\psi\left(\frac{p-i}{b}\right)\\
\nonumber & +\frac{2\zeta_{2}v^{2}}{Nb}\sum_{|p|\le{n}}{\bf
E}\{G_{1}^{2}(i,p)G_{2}(i,p)\}\psi\left(\frac{p-i}{b}\right)\\
\nonumber
&-\frac{\zeta_{2}}{6}\sum_{|p|\le{n}}\frac{\Delta_{pi}}{b^{2}}{\bf
E}\left\{D^{3}_{pi}\left(g^{0}_{1}G_{2}(i,p)\right)\right\}
-\frac{\zeta_{2}\Delta_{ii}}{2b^{2}}{\bf
E}\left\{D^{3}_{ii}\left(g^{0}_{1}G_{2}(i,i)\right)\right\}\\
\label{b.3R12form1}
&-\frac{\zeta_{2}}{120}\sum_{|p|\le{n}}\frac{\theta_{pi}}{b^{3}}(1+\delta_{pi})^{3}{\bf
E}\left\{D^{5}_{pi}\left(g^{0}_{1}G_{2}(i,p)\right)\right\}+\epsilon_{i}
\end{align}
with
\begin{align}
\nonumber \epsilon_{i}= &-\zeta_{2}\sum_{|p|\le{n}}\frac{1}{6!}{\bf
E}\left\{H(p,i)^{7}[D^{6}_{pi}(g^{0}_{1}G_{2}(i,p))]^{(0)}\right\}\\
\nonumber &+\zeta_{2}\sum_{|p|\le{n}} \frac{K_{2}}{5!}{\bf
E}\left\{H(p,i)^{5}[D^{6}_{pi}(g^{0}_{1}G_{2}(i,p))]^{(1)}\right\}\\
\nonumber &+\zeta_{2}\sum_{|p|\le{n}}\frac{K_{4}}{(3!)^{2}}{\bf
E}\left\{H(p,i)^{3}[D^{6}_{pi}(g^{0}_{1}G_{2}(i,p))]^{(2)}\right\}\\
\label{b.3epsiloni} &+\zeta_{2}\sum_{|p|\le{n}}\frac{K_{6}}{5!}{\bf
E}\left\{H(p,i)[D^{6}_{pi}(g^{0}_{1}G_{2}(i,p))]^{(3)}\right\},
\end{align}
where $K_{r}$ are the cumulants of $H(p,i)$ as in
(\ref{b.3K2K4})-(\ref{b.3K6}).

\subsection{Main relation for $R_{12}(i)$}
To give the complete description of $R_{12}$, we use the notation
\begin{equation*}
U(p,i)=\frac{1}{b}\psi\left(\frac{p-i}{b}\right), \quad
U_{G}(i)=\sum_{|p|\le{n}}G(p,p)U(p,i)
\end{equation*}
and introduce the identity
\begin{equation}\label{b.3Efg}
{\bf E}\{\xi g\}={\bf E}\{\xi\}{\bf E}\{g\}+{\bf E}\{\xi g^{0}\}.
\end{equation}
Then, we rewrite the first term of the RHS of (\ref{b.3R12form1}) in
the form
\begin{align*}
\zeta_{2}v^{2}{\bf
E}& \left\{g^{0}_{1}G_{2}(i,i)\sum_{|p|\le{n}}G_{2}(p,p)U(p,i)\right\}\\
&=\zeta_{2}v^{2}R_{12}(i){\bf E}\{U_{G_{2}}(i)\}+\zeta_{2}v^{2}{\bf
E}\{g^{0}_{1}G_{2}(i,i)U^{0}_{G_{2}}(i)\}.
\end{align*}
Now computing the partial derivatives with the help of
(\ref{b.2partialG}), we obtain the following relation for $R_{12}$
\begin{align}
\nonumber R_{12}(i)= &\zeta_{2}v^{2}R_{12}(i){\bf
E}\{U_{G_{2}}(i)\}+\zeta_{2}v^{2}{\bf
E}\{g^{0}_{1}G_{2}(i,i)U^{0}_{G_{2}}(i)\} \\
\label{b.3R12form2} &+\frac{2\zeta_{2}v^{2}}{N}
\sum_{|p|\le{n}}F_{12}(i,p)U(p,i)+\frac{1}{Nb}\Upsilon_{12}(i)
+\sum_{r=1}^{7}Y_{r}(i)+\epsilon_{i}
\end{align}
with $F_{12}(i,p)={\bf E}\{G_{1}^{2}(i,p)G_{2}(i,p)\}$,
\begin{equation}\label{b.3Upsilon}
\Upsilon_{12}(i)=\frac{\zeta_{2}}{b}\sum_{|p|\le{n}}\Delta_{pi}{\bf
E}\left\{[G_{1}^{2}(i,i)G_{1}(p,p)+G_{1}^{2}(p,p)G_{1}(i,i)]
G_{2}(i,i)G_{2}(p,p)\right\},
\end{equation}
the terms $Y_{r}(i)$, $r=1,\ldots,7$ are given by relations
\begin{align*}
Y_{1}(i)=&\frac{\zeta_{2}}{b^{2}}\sum_{|p|\le{n}}{\bf
E}\{g^{0}_{1}G_{2}(i,i)^{2}G_{2}(p,p)^{2}\}\Delta_{pi},\\
Y_{2}(i)=&\zeta_{2}v^{2}{\bf
E}\left\{g^{0}_{1}\sum_{|p|\le{n}}G_{2}(i,p)^{2}U(p,i)\right\},\\
Y_{3}(i)=&\frac{\zeta_{2}}{b^{2}}\sum_{|p|\le{n}}{\bf
E}\left\{g^{0}_{1}G_{2}(i,p)^{4}
+6g^{0}_{1}G_{2}(i,p)^{2}G_{2}(i,i)G_{2}(p,p)\right\}\Delta_{pi},\\
Y_{4}(i)=&\frac{2\zeta_{2}}{Nb^{2}}\sum_{|p|\le{n}}{\bf
E}\left\{G_{1}^{2}(i,p)G_{2}(i,p)^{3}
+3G_{1}^{2}(i,p)G_{2}(i,p)G_{2}(i,i)G_{2}(p,p)\right\}\Delta_{pi},
\end{align*}

\begin{align*}
Y_{5}(i)= &\frac{2\zeta_{2}}{Nb^{2}}\sum_{|p|\le{n}}{\bf
E}\left\{G_{1}^{2}(i,p)G_{1}(i,p)G_{2}(i,p)^{2}
+G_{1}^{2}(i,p)G_{1}(i,p)G_{2}(i,i)G_{2}(p,p)\right\}\Delta_{pi}\\
 &+\frac{\zeta_{2}}{Nb^{2}}\sum_{|p|\le{n}}{\bf
E}\left\{G_{1}^{2}(i,i)G_{1}(p,p)G_{2}(i,p)^{2}
+G_{1}^{2}(p,p)G_{1}(i,i)G_{2}(i,p)^{2}\right\}\Delta_{pi}\\
&+\frac{2\zeta_{2}}{Nb^{2}}\sum_{|p|\le{n}}{\bf
E}\left\{G_{1}^{2}(i,p)G_{1}(i,p)^{2}G_{2}(i,p)
+G_{1}^{2}(i,i)G_{1}(p,p)G_{1}(i,p)G_{2}(i,p)\right\}\Delta_{pi}\\
&+\frac{2\zeta_{2}}{Nb^{2}}\sum_{|p|\le{n}}{\bf
E}\left\{G_{1}^{2}(i,p)G_{1}(p,p)G_{1}(i,i)G_{2}(i,p)\right\}\Delta_{pi}\\
&+\frac{2\zeta_{2}}{Nb^{2}}\sum_{|p|\le{n}}{\bf
E}\left\{G_{1}^{2}(p,p)G_{1}(i,i)G_{1}(i,p)G_{2}(i,p)\right\}\Delta_{pi},\\
Y_{6}(i)= &-\frac{3\zeta_{2}}{b^{2}}\left({\bf
E}\{g_{1}^{0}G_{2}(i,i)^{4}\}+\frac{1}{N}{\bf
E}\{G^{2}_{1}(i,i)G_{2}(i,i)^{3}\}\right)\Delta_{ii}\\
& -\frac{3\zeta_{2}}{Nb^{2}}{\bf
E}\left\{G^{2}_{1}(i,i)G_{1}(i,i)G_{2}(i,i)[G_{1}(i,i)+G_{2}(i,i)]
\right\}\Delta_{ii},\\
Y_{7}(i)=
&-\frac{\zeta_{2}}{120}\sum_{|p|\le{n}}\frac{\theta_{pi}}{b^{3}}
(1+\delta_{pi})^{3}{\bf
E}\left\{D^{5}_{pi}(g^{0}_{1}G_{2}(i,p))\right\}
\end{align*}
and $\epsilon_{i}$ given by (\ref{b.3epsiloni}). The first and the
second terms of the RHS of (\ref{b.3R12form2}) is expressed in terms
of $R_{12}$ and this finally gives a closed relation for $R_{12}$.
The third and forth terms of the RHS of (\ref{b.3R12form2}) give a
non-zero contribution to $R_{12}$ that provide the expression of the
leading term $T(z_{1},z_{2})$ (\ref{b.2Tz1z2}). We will compute this
contribution later (see subsection 4.3). The two last terms of
(\ref{b.3R12form2}) contributes with $o((Nb)^{-1})$ to
(\ref{b.2Cnbz1z2Th}). We formalize this proposition in the following
two statements. \vskip0,2cm

\begin{lem} Under conditions of Theorem 2.1, the estimate
\begin{equation}\label{b.3Y1Y2lem}
\max_{r=1,2}\left\{\sup_{|i|\le{n}} |Y_{r}(i)|\right\}
=O\left(b^{-2}n^{-1}+b^{-2}[{\bf Var}\{g_{1}\}]^{1/2}\right).
\end{equation}
is true in the limit $n,b\rightarrow\infty$ (\ref{b.2bnalpha}).
\end{lem}\vskip0,2cm
We postpone the proof of Lemma 3.1 to the next section. \vskip0,2cm

\begin{lem} Under conditions of Theorem 2.1, the estimate
\begin{equation}\label{b.3Y3Y6lem}
\max_{r=3,4,5,6,7}\left\{\sup_{|i|\le{n}} |Y_{r}(i)|\right\}
=O\left(b^{-2}n^{-1}+b^{-2}[{\bf Var}\{g_{1}\}]^{1/2}\right)
\end{equation}
and
\begin{equation}\label{b.3epsilonlem}
\sup_{|i|\le{n}}|\epsilon_{i}|=O\left(b^{-2}n^{-1}+b^{-2}[{\bf
Var}\{g_{1}\}]^{1/2}\right)
\end{equation}
are true in the limit $n,b\rightarrow\infty$ (\ref{b.2bnalpha}).
\end{lem}\vskip0,2cm

{\it Proof of Lemma 3.2.} We start with (\ref{b.3Y3Y6lem}).
Inequality (\ref{b.3GijleImz}) and (\ref{b.3sumGij2leImz2}) implies
that if $z_{l}\in \Lambda_{\eta}$, then
\begin{equation*}
|Y_{3}(i)|\le{\frac{7[V_{4}+3v^{4}]}{\eta^{3}b^{2}}\sum_{|p|\le{n}}{\bf
E}|g_{1}^{0}G_{2}(i,p)^{2}|}=O\left(\frac{1}{b^{2}}\{{\bf
Var}\{g_{1}\}\}^{1/2}\right).
\end{equation*}
To estimate $Y_{r}$, $r=4,5$, we use (\ref{b.3GijleImz}),
(\ref{b.3sumGij2leImz2}) and inequality
\begin{align}
\nonumber \sum_{|i|\le{n}}{\bf E}|G^{m}_{1}(i,p)G_{2}(i,p)|&\le{{\bf
E}\left(\sum_{|i|\le{n}}|G^{m}_{1}(i,p)|^{2}\right)^ {1/2}
\left(\sum_{|i|\le{n}}|G_{2}(i,p)|^{2}\right)^{1/2}}\\
\label{b.3sumG1ipG2ip}&\le{\frac{1}{\eta^{m+1}}}
\end{align}
with $m=1,2$. Then we get that
$|Y_{4}(i)|\le{8[V_{4}+v^{4}]/(\eta^{6}Nb^{2})}$. Using
(\ref{b.3GijleImz}), (\ref{b.3sumGij2leImz2}) and
(\ref{b.3sumG1ipG2ip}) with $m=1,2$, we obtain that the terms
$\sup_{i}|Y_{r}(i)|$, $r=5,6$ are all of the order indicated in
(\ref{b.3Y3Y6lem}).\\
Let us estimate $Y_{7}$. Let us accept for the moment that
\begin{equation}\label{b.3ED5pig10G2ij}
{\bf E}|D^{5}_{pi}\{g^{0}_{1}G_{2}(i,p)\}|=O\left(N^{-1}+[{\bf
Var}\{g_{1}\}]^{1/2}\right), \ \hbox{ as } \quad
n,p\rightarrow\infty
\end{equation}
holds. Using this estimate and relation (\ref{b.2psicondit}), we
obtain that
\begin{equation*}
\sum_{|p|\le{n}}\left|\frac{\theta_{pi}}{b}\right|\le{c\sum_{|p|\le{n}}\frac{1}{b}\psi\left(\frac{p-i}{b}\right)}=O(1)
\end{equation*}
and that
\begin{equation*}
\sup_{|i|\le{n}} |Y_{7}(i)|=O\left(b^{-2}n^{-1}+b^{-2}\{{\bf
Var}\{g_{1}\}\}^{1/2}\right)
\end{equation*}
where $c$ is a constant.

Now let use prove (\ref{b.3ED5pig10G2ij}). Using (\ref{b.2partialG})
and (\ref{b.3GijleImz}), we get for $z_{1}\in\Lambda_{\eta}$
\begin{equation*}
D_{pi}\{g^{0}_{1}\}=\frac{1}{N}\sum_{|t|\le{n}}D_{pi}\{G_{1}(t,t)\}
=-\frac{2}{N}G^{2}_{1}(i,p)=O\left(\frac{1}{N}\right).
\end{equation*}
It is easy to show that
\begin{equation}\label{b.3Drpig10}
D^{r}_{pi}\{g^{0}_{1}\}=O\left(\frac{1}{N}\right), \quad
r=1,2,\ldots, \ z\in\Lambda_{\eta}.
\end{equation}
Then (\ref{b.3ED5pig10G2ij}) follows from (\ref{b.3Drpig10}) and
(\ref{b.3GijleImz}). Estimate (\ref{b.3Y3Y6lem}) is
proved.\vskip0,5cm

To proceed with estimates of  $\epsilon_{i}$ (\ref{b.3epsilonlem}),
we use the following simple statement, proved in the previous work
\cite{A}. \vskip0,2cm

\begin{lem} (see \cite{A}) If $z_{l}\in\Lambda_{\eta},
\ l=1,2$, under conditions of Theorem 2.1, the estimates
\begin{equation}\label{b.3Vargnbnu}
{\bf Var}([g_{n,b}(z_{l})]^{(\nu)})=O\left({\bf
Var}\{g_{n,b}(z_{l})\}+b^{-1}N^{-2}\right), \quad \nu=0,\ldots,3
\end{equation}
and
\begin{equation}\label{b.3D6pig10G2ip}
D^{6}_{pi}\left\{g^{0}_{1}G_{2}(i,p)\right\}=O\left(N^{-1}+|g_{1}^{0}|\right)
\end{equation}
are true in the limit  $n,b\longrightarrow \infty$
(\ref{b.2bnalpha}).
\end{lem}
\vskip0,2cm

Now regarding the first term of (\ref{b.3epsiloni}) and using
(\ref{b.3Vargnbnu}) and (\ref{b.3D6pig10G2ip}), we obtain inequality
\begin{align}
\nonumber &\sum_{|p|\le{n}}{\bf
E}|H(p,i)^{7}[D^{6}_{pi}(g^{0}_{1}G_{2}(i,p))]^{(0)}|\le{c_{1}\sum_{|p|\le{n}}{\bf
E}\left\{\frac{|H(p,i)|^{7}}{N}+|H(p,i)|^{7}|[g^{0}_{1}]^{(0)}|\right\}}\\
\nonumber &\le{c_{1}\sum_{|p|\le{n}}\frac{\hat{\mu}_{7}}{Nb^{7/2}}
\psi\left(\frac{p-i}{b}\right)+c_{1}\sum_{|p|\le{n}}\frac{(\hat{\mu}_{14})^{1/2}}{b^{7/2}}
\left(\psi\left(\frac{p-i}{b}\right)\right)^{1/2}\left({\bf
Var}\{[g_{1}]^{(0)}\}\right)^{1/2}} \\
\label{b.3epsilonestim1} &=O\left(N^{-1}b^{-2}+b^{-2}[{\bf
Var}\{g_{1}\}]^{1/2}\right),
\end{align}
where $c$ is a constant.

 Regarding the last term of the right-hand side of (\ref{b.3epsiloni})
and using (\ref{b.3Vargnbnu}) and (\ref{b.3D6pig10G2ip}), we obtain
inequality
\begin{align}
\nonumber &\sum_{|p|\le{n}}K_{6}{\bf
E}|H(p,i)[D^{6}_{pi}(g^{0}_{1}G_{2}(i,p))]^{(3)}|\\
\nonumber
&\le{\frac{c_{2}}{b^{2}}\left(\frac{1}{b}\sum_{|p|\le{n}}\psi\left(\frac{p-i}{b}\right)\right)
\left(\frac{\hat{\mu}_{1}\psi(\frac{p-i}{b})}{Nb^{1/2}}+\frac{\hat{\mu}^{1/2}_{2}
\left(\psi(\frac{p-i}{b})\right)^{1/2}}{b^{1/2}}\left({\bf
Var}\{[g_{1}]^{(3)}\}\right)^{1/2}\right)} \\
\label{b.3epsilonestim2} &=O\left(N^{-1}b^{-2}+b^{-2}[{\bf
Var}\{g_{1}\}]^{1/2}\right),
\end{align}
where $c_{2}$ is a constant.

 Repeating previous computations of
(\ref{b.3epsilonestim2}), we obtain that
\begin{align}
\nonumber &\sum_{|p|\le{n}}K_{4}{\bf
E}|H(p,i)^{3}[D^{6}_{pi}(g^{0}_{1}G_{2}(i,p))]^{(2)}|
+\sum_{|p|\le{n}}K_{2}{\bf
E}|H(p,i)^{5}D^{6}_{pi}[g^{0}_{1}G_{2}(i,p)]^{(1)}|\\
\label{b.3epsilonestim3} &=O\left(N^{-1}b^{-2}+b^{-2}[{\bf
Var}\{g_{1}\}]^{1/2}\right).
\end{align}
Then (\ref{b.3epsilonlem}) follows from the estimates given by
relations (\ref{b.3epsilonestim1}), (\ref{b.3epsilonestim2}) and
(\ref{b.3epsilonestim3}). Lemma 3.2 is proved. $\hfill \blacksquare$

\vskip0,2cm Let us come back to relation (\ref{b.3R12form2}). Using
equality
\begin{equation*}
\sum_{|p|\le{n}}{\bf E}\{g^{0}_{1}G_{2}(i,i)G^{0}_{2}(p,p)\}U(p,i) =
{\bf E}\{g^{0}_{1}U_{G_{2}}^{0}(i)G_{2}^{0}(i,i)\} +
U_{R_{12}}(i){\bf E}\{G_{2}(i,i)\},
\end{equation*}
we obtain the following relation
\begin{align}
\nonumber R_{12}(i) = &\zeta_{2}v^{2}R_{12}(i)U_{{\bf
E}(G_{2})}(i)+\zeta_{2}v^{2}U_{R_{12}}(i){\bf E}\{G_{2}(i,i)\}\\
\nonumber
&+\frac{2\zeta_{2}v^{2}}{N}\sum_{|p|\le{n}}F_{12}(i,p)U(p,i)+
\frac{1}{Nb}\Upsilon_{12}(i)\\
\label{b.3R12form3} &+\tau(i)+\sum_{r=1}^{7}Y_{r}(i)+\epsilon_{i},
\end{align}
where $F_{12}(i,p)$ is the same as in (\ref{b.3R12form2}),
$\Upsilon_{12}$ is given by (\ref{b.3Upsilon}) and
\begin{equation}\label{b.3tau}
\tau(i) = \zeta_{2}v^{2}{\bf
E}\{g^{0}_{1}U_{G_{2}}^{0}(i)G_{2}^{0}(i,i) \}.
\end{equation}
Relation (\ref{b.3R12form3}) is the main equality used for the proof
of Theorem 2.1. We use (\ref{b.3R12form3}) twice : at the first
stage we estimate the variance ${\bf Var}\{g_{n,b}(z)\}$ and at the
second one we obtain explicit expressions for the leading term of
$C_{n,b}(z_{1},z_{2})$. This will be done this in the next section.
\vskip0,5cm

\section{Variance and leading term of $C_{n,b}(z_{1},z_{2})$}
In this section we give the estimate of the variance and the proof
of Theorem 2.1, postponing some technical results to the next
section. \vskip0,5cm

\subsection{Estimate of the variance}
Let us define an auxiliary variable
\begin{equation}\label{b.4q2i}
q_{2}(i)=\frac{\zeta_{2}}{1-\zeta_{2}v^{2}U_{g_{2}}(i)},
\end{equation}
where $g_{2}(i)={\bf E}G_{2}(i,i)$. Then we can rewrite
(\ref{b.3R12form3}) in the form
\begin{align}
\nonumber R_{12}(i)= &
v^{2}q_{2}(i)U_{R_{12}}(i)g_{2}(i)+\frac{1}{Nb}
\left(2v^{2}q_{2}(i)b[F_{12}U](i,i)
+q_{2}(i)\zeta^{-1}_{2}\Upsilon_{12}(i)\right) \\
\label{b.4R12form4}
&+q_{2}(i)\zeta^{-1}_{2}\left(\tau(i)+\sum_{r=1}^{7}Y_{r}(i)
+\epsilon_{i}\right)
\end{align}
with
\begin{equation*}
[F_{12}U](i,i)=\sum_{|p|\le{n}}F_{12}(i,p)U(p,i),
\end{equation*}
where $F_{12}(i,p)$ is the same as in (\ref{b.3R12form2}) and
$\Upsilon_{12}$ is given by (\ref{b.3Upsilon}).

Now let us estimate the terms of the RHS of
(\ref{b.4R12form4}).Taking into account that $U(p,i)\le{b^{-1}}$ and
using inequalities (\ref{b.3sumG1ipG2ip}) with $m=2$, it is easy to
see that if $z_{l}\in\Lambda_{\eta}$, then
\begin{equation}\label{b.4estim1}
\frac{1}{N}|[F_{12}U](i,i)|\le{\frac{1}{\eta^{3}Nb}}
=O(\frac{1}{Nb}).
\end{equation}
Let us estimate $\Upsilon_{12}$ (\ref{b.3Upsilon}). Using
(\ref{b.3GijleImz}) and inequality
$|\Delta_{pi}|\le{[V_{4}+3v^{4}]\psi((p-i)/b)}$, we obtain that
\begin{equation}\label{b.4q2ileImz}
|q_{2}(i)|\le{\frac{1}{|\Im z_{2}|}}, \quad z_{2}\in\Lambda_{\eta}
\end{equation}
and that
\begin{equation}\label{b.4estim2}
|q_{2}(i)\zeta^{-1}_{2}\Upsilon_{12}(i)|\le{\frac{2[V_{4}+3v^{4}]}{\eta^{6}}
\sum_{|p|\le{n}}\frac{1}{b}\psi\left(\frac{p-i}{b}\right)}=O(1).
\end{equation}
To estimate the term $\tau$ (\ref{b.3tau}), we use the following
statement. \vskip0,2cm
\begin{lem} Under the conditions of Theorem 2.1, the estimate
\begin{equation}\label{b.4taulem}
\sup_{|i|,|s|\le{n}}|{\bf E}g^{0}(z)G^{0}(i,i)U^{0}_{G}(s)| =
O\left(n^{-1}b^{-2}+b^{-2}[{\bf Var}\{g(z)\}]^{1/2}\right)
\end{equation}
is true in the limit $n,b\rightarrow\infty$ (\ref{b.2bnalpha}).
\end{lem}
\vskip0,2cm We prove Lemma 4.1 in section 5. \vskip0,2cm

It follows from results of Lemmas 3.1, 3.2 and relation
(\ref{b.4taulem}), that if $z_{j}\in \Lambda_{\eta}$, then
\begin{equation}\label{b.4estim3}
\sup_{|i|\le{n}}\left|q_{2}(i)\zeta^{-1}_{2}\left(\tau(i)+
\sum_{r=1}^{7}Y_{r}(i)+\epsilon_{i}\right)\right|
=O\left(N^{-1}b^{-2}+b^{-2}\{{\bf Var}\{g_{1}\}\}^{1/2}\right).
\end{equation}
Let us denote $r_{12} = \sup_{i}|R_{12}(i)|$. Regarding estimates
(\ref{b.4estim1}), (\ref{b.4estim2}) and (\ref{b.4estim3}), we
derive from (\ref{b.4R12form4}) inequality
\begin{equation*}
r_{12}\le{ \frac{v^{2}}{\eta^{2}}r_{12} + \frac{A}{bN} +
\frac{1}{b^{2}}\sqrt{r_{12}}}
\end{equation*}
for some constant A. Since $r_{12}$ is bounded for all $z_{l}\in
\Lambda_{\eta}$, then $r_{12}=O((Nb)^{-1}+b^{-4})$. Using condition
(\ref{b.2bnalpha}) and taking $z=z_{1}=\overline{z_{2}}$, one
obtains that
\begin{equation}\label{b.4Vargnbz}
{\bf Var}\{g_{n,b}(z)\}=O\left(\frac{1}{Nb}\right).
\end{equation}
Substituting (\ref{b.4Vargnbz}) into (\ref{b.4estim3}), we obtain
that
\begin{equation}\label{b.4estim5}
\sup_{|i|\le{n}}\left|q_{2}(i)\zeta^{-1}_{2}\left(\tau(i)+
\sum_{r=1}^{7}Y_{r}(i)+\epsilon_{i}\right)\right|
=o\left(\frac{1}{Nb}\right)
\end{equation}
in the limit $n,b\rightarrow\infty$ (\ref{b.2bnalpha}) and for all
$z_{l}\in\Lambda_{\eta}$, $l=1,2$. This proves (\ref{b.2Cnbz1z2Th}).
\vskip0,5cm

\subsection{Leading term of the correlation function}
Assuming that (\ref{b.4estim5}) is true, we rewrite
(\ref{b.4R12form4}) in the form
\begin{equation}\label{b.4R12form5}
R_{12}(i)=v^{2}q_{2}(i)g_{2}(i)U_{R_{12}}(i)+\frac{1}{Nb}f_{12}(i)
+\Gamma(i)
\end{equation}
with
\begin{equation}\label{b.4f12}
f_{12}(i)=
2v^{2}q_{2}(i)b[F_{12}U](i,i)+q_{2}(i)\zeta^{-1}_{2}\Upsilon_{12}(i),
\end{equation}
where $F_{12}(i,p)$ is the same as in (\ref{b.3R12form3}) and
$\Upsilon_{12}$ is given by (\ref{b.3Upsilon}). We have denoted the
vanishing terms by
\begin{equation}\label{b.4Gamma}
\Gamma(i)=q_{2}(i)\zeta^{-1}_{2}\left(\tau(i)+
\sum_{r=1}^{7}Y_{r}(i)+\epsilon_{i}\right).
\end{equation}
To obtain an explicit expression for the leading term of
$C_{n,b}(z_{1},z_{2})$, it is necessary to study in detail the
variables $F_{12}$ and $\Upsilon_{12}$. Let us formulate the
corresponding statements and the auxiliary relations needed. Given a
positive integer $L$, set
\begin{equation}\label{b.4BL}
B_{L}\equiv B_{L}(n,b)=\left\{i\in \Z ; \ |i|\le{n-bL}\right\}.
\end{equation}

\vskip0,2cm
\begin{lem} If  $z\in\Lambda_{\eta}$, then for arbitrary
positive $\epsilon$ and large enough values of $n$ and $b$
(\ref{b.2bnalpha}) there exists a positive integer $L=L(\epsilon)$
such that relations
\begin{equation}\label{b.4supbF12U}
\sup_{i\in{B_{L}}}\left|b[F_{12}U](i,i)
-\frac{w_{2}w_{1}^{2}}{2\pi(1- v^{2}w_{1}^{2})}\int_{\R}
\frac{\tilde{\psi}(p)}{[1-
v^{2}w_{1}w_{2}\tilde{\psi}(p)]^{2}}dp\right|\le{\epsilon}
\end{equation}
and
\begin{equation}\label{b.4supUpsilon}
\sup_{i\in{B_{L}}}\left|q_{2}(i)\zeta^{-1}_{2}\Upsilon_{12}(i)-
\frac{2\Delta
w^{3}_{1}w^{3}_{2}}{1-v^{2}w^{2}_{1}}\right|\le{\epsilon}
\end{equation}
hold for enough $n$ and $b$ satisfying (\ref{b.2bnalpha}) with
$\Delta$ is given by (\ref{b.2Delta}).
\end{lem}
\vskip0,2cm

The proof of Lemma 4.2 is based on the following statement
formulated for the product $G_{1}G_{2}$. \vskip0,2cm

\begin{lem} Given positive $\epsilon$, there exists a positive integer
$L=L(\epsilon)$ such that relations
\begin{equation}\label{b.4supEG2ii}
\sup_{i\in{B_{L}}}\left|{\bf E}\{G^{2}_{1}(i,i)\} -
\frac{w^{2}_{1}}{1 - v^{2}w^{2}_{1}}\right|\le{\epsilon},
\end{equation}
\begin{equation}\label{b.4supbET12U}
\sup_{i\in{B_{L}}}\left|b\sum_{|s|\le{n}}{\bf
E}\{G_{1}(i,s)G_{2}(i,s)\}U^{k}(s,i)- \frac{1}{2\pi}\int_{\R}
\frac{w_{1}w_{2}\tilde{\psi}^{k}(p)}{1-
v^{2}w_{1}w_{2}\tilde{\psi}(p)}dp\right|\le{\epsilon}
\end{equation}
and
\begin{equation}\label{b.4supET12}
\sup_{i\in{B_{L}}}\left|\sum_{|s|\le{n}}{\bf
E}\{G_{1}(i,s)G_{2}(i,s)\} - \frac{w_{1}w_{2}}{1 -
v^{2}w_{1}w_{2}}\right|\le{\epsilon}
\end{equation}
hold for enough $n$ and $b$ satisfying (\ref{b.2bnalpha}) for all
$k\in \N$, all $z_{j}\in \Lambda_{\eta}$, $j=1,2$.
\end{lem}
\vskip0,2cm We postpone the proof of Lemma 4.3 to the next section.
\vskip0,2cm

\subsection{Proof of Lemma 4.2 and Theorem 2.1}
\subsubsection{Proof of Lemma 4.2}
We start with (\ref{b.4supbF12U}). Let us consider the average
$F_{12}(i,s)={\bf E}\{G^{2}_{1}(i,s)G_{2}(i,s)\}$. Applying to
$G_{2}(i,s)$ the resolvent identity (\ref{b.2h-zI-1}), we obtain
equality
\begin{equation*}
F_{12}(i,s)=\zeta_{2}\delta_{is}{\bf
E}\{G^{2}_{1}(i,i)\}-\zeta_{2}\sum_{|p|\le{n}}{\bf
E}\{G^{2}_{1}(i,s)G_{2}(i,p)H(p,s)\}.
\end{equation*}
Applying formula (\ref{b.2EXjFX1Xm}) to ${\bf
E}\{G^{2}_{1}(i,s)G_{2}(i,p)H(p,s)\}$ with $q=3$ and taking into
account (\ref{b.2partialG}), we get relation
\begin{align}
\nonumber F_{12}(i,s)= &\zeta_{2}\delta_{is}{\bf
E}\{G^{2}_{1}(i,i)\}+ \zeta_{2}v^{2}[t_{12}U](i,s){\bf
E}\{G^{2}_{1}(s,s)\}\\
\nonumber &+\zeta_{2}v^{2}[F_{12}U](i,s)g_{1}(s)+
\zeta_{2}v^{2}F_{12}(i,s)U_{g_{2}}(s)\\
\label{b.4F12} &+\sum_{r=1}^{5}\beta_{r}(i,s),
\end{align}
where we denoted $g_{l}(s)={\bf E}\{G_{l}(s,s)\}$, $l=1,2$,
$t_{12}(i,s)={\bf E}\{G_{1}(i,s)G_{2}(i,s)\}$ and the terms
$\beta_{l}$, $l=1,\ldots,5$ are given by:
\begin{align*}
\beta_{1}(i,s)=&\zeta_{2}v^{2}\sum_{|p|\le{n}}{\bf
E}\{G^{2}_{1}(p,s)G_{1}(i,s)G_{2}(i,p)\}U(p,s)\\
&+\zeta_{2}v^{2}\sum_{|p|\le{n}}{\bf
E}\{G^{2}_{1}(i,s)G_{1}(p,s)G_{2}(i,p)\}U(p,s)\\
&+\zeta_{2}v^{2}\sum_{|p|\le{n}}{\bf
E}\{G^{2}_{1}(i,s)G_{2}(p,s)G_{2}(i,p)\}U(p,s),\\
\beta_{2}(i,s)=&\zeta_{2}v^{2}\sum_{|p|\le{n}}{\bf
E}\{G_{1}(i,p)G_{2}(i,p)(G^{2}_{1}(s,s))^{0}\}U(p,s)\\
&+\zeta_{2}v^{2}\sum_{|p|\le{n}}{\bf
E}\{G^{2}_{1}(i,p)G_{2}(i,p)G^{0}_{1}(s,s)\}U(p,s),\\
\beta_{3}(i,s)= &\zeta_{2}v^{2}{\bf
E}\left\{G^{2}_{1}(i,s)G_{2}(i,s)U^{0}_{G_{2}}(s)\right\},\\
\beta_{4}(i,s)= &-\frac{\zeta_{2}}{6}\sum_{|p|\le{n}}K_{4}{\bf
E}\left\{D^{3}_{ps}\left(G^{2}(i,s)G_{2}(i,p)\right)\right\},
\end{align*}
and
\begin{align*}
\beta_{5}(i,s)=&-\frac{\zeta_{2}}{4!}\sum_{|p|\le{n}}{\bf
E}\left\{H(p,s)^{5}[D^{4}_{ps}(G^{2}(i,s)G_{2}(i,p))]^{(0)}\right\}\\
&+\frac{\zeta_{2}}{3!}\sum_{|p|\le{n}}K_{2}{\bf
E}\left\{H(p,s)^{3}[D^{4}_{ps}(G^{2}(i,s)G_{2}(i,p))]^{(1)}\right\}\\
&+\frac{\zeta_{2}}{3!}\sum_{|p|\le{n}}K_{4}{\bf
E}\left\{H(p,s)[D^{4}_{ps}(G^{2}(i,s)G_{2}(i,p))]^{(2)}\right\}
\end{align*}
with $K_{r}$ are the cumulants of $H(p,s)$ as in
(\ref{b.3K2K4})-(\ref{b.3K6}).

Let us accept for the moment that
\begin{equation}\label{b.4betaresti}
\max_{j=1,\ldots,5}\left\{\sup_{|i|,|s|\le{n}}|
\beta_{r}(i,s)|\right\}=O\left(b^{-1}\right), \quad \hbox{ as }
\quad n,b\rightarrow\infty
\end{equation}
holds for enough $n$ and $b$ satisfying (\ref{b.2bnalpha}). Using
them and the definition of $q_{2}(s)$ (\ref{b.4q2i}), we rewrite
(\ref{b.4F12}) in the form
\begin{equation}\label{b.4F12form1}
F_{12}(i,s)=
v^{2}g_{1}(s)q_{2}(s)[F_{12}U](i,s)+R_{1}(i,s)+R_{2}(i,s)+\beta(i,s),
\end{equation}
where we denoted
\begin{equation}\label{b.4R1}
R_{1}(i,s)=q_{2}(i){\bf E}\{G^{2}_{1}(i,i)\}\delta_{is},
\end{equation}
\begin{equation}\label{b.4R2}
R_{2}(i,s)=v^{2}q_{2}(s)[t_{12}U](i,s){\bf E}\{G^{2}_{1}(s,s)\}
\end{equation}
and the vanishing term
\begin{equation*}
\beta(i,s)=\frac{q_{2}(s)}{\zeta_{2}}\sum_{r=1}^{5}\beta_{r}(i,s).
\end{equation*}
We define the linear operator $W$ that acts on the space of
$N\times{N}$ matrices $F$ according to the formula
\begin{equation*}
[WF](i,s)=v^{2}g_{1}(s)q_{2}(s)\sum_{|p|\le{n}}F(i,p)U(p,s).
\end{equation*}
It is easy to see that if $z_{l}\in\Lambda_{\eta}$, then the
estimates (\ref{b.3GijleImz}) and (\ref{b.4q2ileImz}) imply that
$|R_{1}|\le{\eta^{-3}} $ and $|R_{2}|\le{v^{2}\eta^{-5}}$ and that
\begin{equation}\label{b.4Winequ}
||W||_{(1,1)}\le{\frac{v^{2}}{\eta^{2}}}<\frac{1}{2},
\end{equation}
where the norm of $N\times{N}$ matrix $A$ is determined as
$||A||_{(1,1)}=\sup_{i,s}|A(i,s)|$. This estimate verified by the
direct computation of the norm $||WA||_{(1,1)}$ with
$||A||_{(1,1)}=1$. Then (\ref{b.4F12form1}) can be rewritten as
\begin{equation}\label{b.4F12form2}
F_{12}(i,s)=\sum_{m=0}^{\infty}\left[W^{m}\left(R_{1} + R_{2}
+\beta\right)\right](i,s).
\end{equation}
The next steps of the proof of (\ref{b.4supbF12U}) are very
elementary. To do this, we start with the following statements,
proved in the previous work \cite{A}. \vskip0,2cm

\begin{lem} (see \cite{A}) Given positive $\epsilon$, there exists a positive integer
$L=L(\epsilon)$ such that relations
\begin{equation}\label{b.4supEGii}
\sup_{i\in{B_{L}}}|{\bf E}\{G(i,i;z)\}-w(z)|\le{\epsilon}, \quad
z\in\Lambda_{\eta}
\end{equation}
and
\begin{equation}\label{b.4supq}
\sup_{i\in{B_{L}}}|q(i;z)-w(z)|\le{2\epsilon} \quad
z\in\Lambda_{\eta}
\end{equation}
hold for enough $n$ and $b$ satisfying (\ref{b.2bnalpha}), where $w$
and $q$ are given by (\ref{b.2wz}) and (\ref{b.4q2i}).
\end{lem}
\vskip0,2cm

Now let us return to relation (\ref{b.4F12form2}). We consider the
first $M$ terms of the infinite series and use the decay of the
matrix elements $U(i,s)=U^{(b)}(i,s)$. If one considers
(\ref{b.4R1}) and (\ref{b.4R2}) with $i$ and $s$ taken far enough
from the endpoints -$n$, $n$, then the variables $g_{1}(j)$,
$q_{2}(k)$ enter into the finite series with $j$ and $k$ also far
from the endpoints. Then one can use relations (\ref{b.4supEGii})
and (\ref{b.4supq}) and replace $g_{1}$ and $q_{2}$ by the constant
values $w_{1}$ and $w_{2}$, respectively. This substitution leads to
simplified expressions with error terms that vanish as
$n,b\rightarrow\infty$. The second step is similar. It is to show
that we can use Lemma 4.3 and replace the terms $R_{1}$ and $R_{2}$
of the finite series of (\ref{b.4R1})and (\ref{b.4R2}) by
corresponding expressions given by formulas (\ref{b.4supEG2ii}) and
(\ref{b.4supbET12U}).

 Let us start to perform this program. Taking into account the estimate of
$\beta$ (\ref{b.4betaresti}) and using bounded-ness of the terms
$R_{1}$ and $R_{2}$, we can deduce from (\ref{b.4F12form2}) equality
\begin{equation}\label{b.4F12form3}
b\sum_{|s|\le{n}}F_{12}(i,s)U(s,i)=b\sum_{m=0}^{M}\left[W^{m}(R_{1}
+ R_{2}).U\right](i,i)+\kappa_{1}(i,i),
\end{equation}
where $ M>0$ is such that given $\epsilon>0$ and
$|\kappa_{1}(i,i)|<\epsilon$ for large enough $b$ and $N$. Now let
us find such $h>0$ that the following holds
\begin{equation*}
\sup_{h\le{|t|}}\psi(t)<\epsilon \ \hbox{ and } \
\int_{h\le{|t|}}\psi(t)dt\le{\epsilon}.
\end{equation*}
We determine the matrix
\begin{equation*}
\hat{U}(i,p)=\left\{
\begin{array}{lll}
U(i,p) & \textrm{if} & |i-p|\le{bh}; \\
0    &  \textrm{if} &  |i-p|>bh
\end{array}\right.
\end{equation*}
and denote by $\hat{W}$ the corresponding linear operator
\begin{equation*}
[\hat{W}F](i,s)=v^{2}g_{1}(s)q_{2}(s)\sum_{|p|\le{n}}F_{12}(i,p)
\hat{U}(p,s).
\end{equation*}
Certainly , $\hat{W}$ admits the same estimate as $W$
(\ref{b.4Winequ}). Given $\epsilon>0$ and $L>0$ the large number.
Let us denote by $Q$ the first natural greater than $(M+k)h$. Then
one can write that
\begin{equation}\label{b.4F12form4}
b\sum_{m=0}^{M}\left[W^{m}(R_{1} +
R_{2}).U\right](i,i)=b\sum_{m=0}^{M}\left[\hat{W}^{m}(R_{1} +
R_{2})\hat{U}\right](i,i) +\kappa_{2}(i,i),
\end{equation}
where
\begin{equation}\label{b.4supkappa2}
\sup_{i\in{B_{L+Q}}}|\kappa_{2}(i,i)|\le{\epsilon} , \quad \hbox{ as
} \  n,b \longrightarrow{\infty}.
\end{equation}
The proof of (\ref{b.4supkappa2}) uses elementary computations.
Indeed, $\kappa_{2}(i,i)$ is represented as the sum of $M+1$ terms
of the form
\begin{equation*}
b\sum_{|s_{r}|\le{n}}^{*}\nu^{2m}g_{1}(s_{1})q_{2}(s_{1})\ldots,g_{1}(s_{m})q_{2}
(s_{m})[R_{1}+R_{2}](i,s_{m+1})
\end{equation*}
\begin{equation*}
\times U(s_{m+1},s_{m})\ldots U(s_{1},i),
\end{equation*}
where the sum is taken over the values of $s_{j}$ such that
$|s_{j}-s_{j+1}|>bh$ at least for one of the numbers $j\le{m}$.

Now remembering the a priori bounds for $R_{1}$ (\ref{b.4R1}) and
$R_{2}$ (\ref{b.4R2}), one obtains the following estimate of
$\kappa_{2}$:
\begin{align}
\nonumber
\sup_{|i|\le{n}}|\kappa_{2}(i,i)|\le&{\sum_{m=0}^{M}\frac{v^{2m}}
{\eta^{2m+3}}\sum_{|s_{r}|\le{n}}^{*}bU(i,s_{1})\ldots
U(s_{m},s_{m+1})} \\
\label{b.4supDelta2estim1}
&+\sum_{m=0}^{M}\frac{v^{2m+2}}{\eta^{2m+5}}
\sum_{|s_{r}|\le{n}}^{*}bU(i,s_{1})\ldots U(s_{m},s_{m+1}).
\end{align}
Assuming that $|s_{j}-s_{j+1}|>bh$ and using inequality
\begin{align}
\sum_{|s_{i}|\le{n}}U(i,s_{1})\ldots U(s_{j-1},s_{j})&\le{
\sum_{s_{i}\in \Z}U(i,s_{1})\ldots U(s_{j-1},s_{j})}\\
\nonumber &\le{\left[\int_{-\infty}^{+\infty}\psi(t)dt
+\frac{\psi(0)}{b}\right]^{j}}\\
\label{b.4sums-r} &\le{(1+1/b)^{j}},
\end{align}
one sees that for large enough $b$ and $n$,
\begin{equation*}
\sum_{|s_{j}|\le{n}}U^{j}(i,s_{j})\epsilon
U^{m-j}(s_{j+1},s_{m+1})\le{\epsilon}.
\end{equation*}
Let us also mention here that given $\epsilon >0$, one has large
enough $n$ and $b$ that
\begin{equation}\label{b.4supUj-1}
\sup_{i\in B_{L+Q}}|\sum_{|s|\le{n}}U^{j}(i,s)-1|\le{\epsilon},
\end{equation}
where $j\le{M}$. This follows from elementary computations related
with the differences
\begin{equation}\label{b.4Pb}
P_{b}=\frac{1}{b}\sum_{t\in
\Z}\psi\left(\frac{t}{b}\right)-\int_{\R}\psi(s)ds
\end{equation}
and
\begin{equation}\label{b.4Tnb}
T_{n,b}(i)\equiv
T(i)=\frac{1}{b}\sum_{|t|\le{n}}\psi\left(\frac{t-i}{b}\right)-
\frac{1}{b}\sum_{t\in \Z}\psi\left(\frac{t}{b}\right).
\end{equation}
that vanish in the limit $1\ll b\ll n$ (see previous work \cite{A}
for more details).

This reasoning when slightly modified is used to estimate the second
term in the RHS of (\ref{b.4supDelta2estim1}). Now one can write
that
\begin{equation*}
\sup_{|i|\le{n}}|\kappa_{2}(i,i)|\le{2\epsilon\sum_{m=0}^{M}m
\left[\frac{v^{2}}{\eta^{2}}\right]^{m}}\le{\epsilon}.
\end{equation*}
Regarding the RHS of (\ref{b.4F12form4}) with $i\in B_{L+Q}$, one
observes that the summations run over such values of $s_{r}$ that
$|i-s_{1}|\le{bh}$, $|s_{r}-s_{r+1}|\le{bh}$, and thus $s_{j}\in
B_{L}$ for all $j\le{k+m-1}$. This means that we can apply relations
(\ref{b.4supEGii}) and (\ref{b.4supq}) to the RHS of
(\ref{b.4F12form4}) and to replace $g_{1}$ by $w_{1}$, $q_{2}$ by
$w_{2}$. From (\ref{b.4F12form3}), it follows that
\begin{align*}
b[F_{12}U](i,i)=&\sum_{m=0}^{M}[v^{2}w_{1}w_{2}]^{m} \
b\sum_{|s_{m+1}|\le{n}}\left(R_{1}(i,s_{m+1})+R_{2}(i,s_{m+1})\right)
\hat{U}^{m+1}(s_{m+1},i)\\
&+\kappa_{3}(i,i)
\end{align*}
with
\begin{equation*}
\sup_{i\in B_{L+Q}}|\kappa_{3}(i,i)|\le{4\epsilon}.
\end{equation*}
Finally, applying Lemma 4.3 to the expressions involved in $R_{l}$
and taking into account that
\begin{equation}\label{b.4bUm+1}
\sup_{i\in
B_{L+Q}}|bU^{m+1}(i,i)-\frac{1}{2\pi}\int_{\R}\tilde{\psi}^
{m+1}(p)dp|\le{\epsilon},
\end{equation}
we obtain equality
\begin{align}
\nonumber
b[F_{12}U](i,i)=&\frac{1}{2\pi}\frac{w^{2}_{1}w_{2}}{1-v^{2}w^{2}_{1}}
\sum_{m=0}^{M}[v^{2}w_{1}w_{2}]^{m}\int_{\R}\tilde{\psi}^{m+1}(p)dp\\
\label{b.4F12form5}
&+\frac{1}{2\pi}\frac{w^{2}_{1}w_{2}}{1-v^{2}w^{2}_{1}}
\sum_{m=0}^{M}[v^{2}w_{1}w_{2}]^{m}v^{2}\int_{\R}
\frac{w_{1}w_{2}\tilde{\psi}^{m+1}(p)}{1-v^{2}w_{1}w_{2}
\tilde{\psi}(p)}dp+ \kappa_{4}(i,i)
\end{align}
with
\begin{equation*}
\sup_{i\in B_{L+Q}}|\kappa_{4}(i,i)|\le{\epsilon}.
\end{equation*}
Passing back in (\ref{b.4F12form5}) to the infinite series and
simplifying them, we arrive at the expression standing in the RHS of
(\ref{b.4supbF12U}). Relation (\ref{b.4supbF12U}) is proved.

\vskip0,5cm

Now let us prove (\ref{b.4betaresti}). Inequality
$U(p,s)\le{b^{-1}}$, (\ref{b.3GijleImz}) and (\ref{b.3sumG1ipG2ip})
imply that if $z_{l}\in\Lambda_{\eta}$, the estimate
\begin{equation}\label{b.4beta12}
\max_{r=1,2}\left\{\sup_{|i|,|s|\le{n}}|\beta_{r}(i,s)|\right\}
=O(b^{-1})
\end{equation}
holds for enough $n$ and $b$ satisfying (\ref{b.2bnalpha}). To
estimate $\beta_{3}$, we use the following estimate of the diagonal
elements of the resolvent $G$, proved in the previous work \cite{A}.
\vskip0,2cm
\begin{lem}
(see \cite{A}) If $z\in\Lambda_{\eta}$, then under conditions of
Theorem 2.1, the estimate
\begin{equation}\label{b.4supEUG02}
\sup_{|s|\le{n}}{\bf E}\{|U^{0}_{G}(s;z)|^{2}\}=O(b^{-2})
\end{equation}
holds for enough $n$ and $b$ satisfying (\ref{b.2bnalpha}).
\end{lem}
\vskip0,2cm

Then inequality (\ref{b.3GijleImz}) and estimate
(\ref{b.4supEUG02}), imply that
\begin{equation}\label{b.4beta3}
\sup_{|i|,|s|\le{n}}|\beta_{3}(i,s)|=O(b^{-1}),\quad
z_{1},z_{2}\in\Lambda_{\eta} \quad \hbox { as } \
n,b\rightarrow\infty.
\end{equation}
Using inequality
\begin{equation}\label{b.4K4estim}
|K_{4}\left(H(p,s)\right)|\le{\frac{4|\Delta_{ps}|}{b^{2}}}
\le{\frac{4[V_{4}+3v^{4}]}{b^{2}}\psi\left(\frac{p-s}{b}\right)}
\end{equation}
and relations (\ref{b.3GijleImz}) and (\ref{b.2partialG}), we obtain
that
\begin{equation*}
|{\bf
E}\left\{D^{3}_{ps}\left(G^{2}(i,s)G_{2}(i,p)\right)\right\}|=O(1),
\quad \hbox { as } \ n,b\rightarrow\infty
\end{equation*}
and conclude that
\begin{equation}\label{b.4beta4}
\sup_{|i|,|s|\le{n}}|\beta_{4}(i,s)|=O(b^{-1}),\quad
z_{1},z_{2}\in\Lambda_{\eta} \quad \hbox { as } \
n,b\rightarrow\infty.
\end{equation}
Regarding the term $\beta_{5}$ and using similar arguments as those
to the proof of (\ref{b.3epsilonlem}) (see
(\ref{b.3epsilonestim1})-(\ref{b.3epsilonestim2})), we conclude that
\begin{equation}\label{b.4beta5}
\sup_{|i|,|s|\le{n}}|\beta_{5}(i,s)|=O(b^{-1}),\quad
z_{1},z_{2}\in\Lambda_{\eta} \quad \hbox { as } \
n,b\rightarrow\infty.
\end{equation}
Now (\ref{b.4betaresti}) follows from (\ref{b.4beta12}),
(\ref{b.4beta3}), (\ref{b.4beta4}) and (\ref{b.4beta5}).

\vskip0,5cm

To complete the proof of Lemma 4.2, let us prove
(\ref{b.4supUpsilon}). To do this we use the following simple
statement, proved in the previous work \cite{A}. \vskip0,2cm
\begin{lem} (see \cite{A}) If $z\in\Lambda_{\eta}$, then under conditions of Theorem 2.1, the
estimate
\begin{equation}\label{b.4supEGss2}
\sup_{|s|\le{n}}{\bf E}\{|G(s,s;z)^{0}|^{2}\}=O(b^{-1})
\end{equation}
holds for enough $n$ and $b$ satisfying (\ref{b.2bnalpha}).
\end{lem}
\vskip0,2cm

We introduce the variable
$M_{12}(i)=q_{2}(i)\zeta^{-1}_{2}\Upsilon_{12}(i)$ with
$\Upsilon_{12}$ is given by (\ref{b.3Upsilon}). Using identity
(\ref{b.3Efg}) and estimate (\ref{b.4supEGss2}), we obtain that
\begin{align}
\nonumber M_{12}(i)= &q_{2}(i)g_{2}(i){\bf
E}\{G^{2}_{1}(i,i)\}\sum_{|p|\le{n}}\frac{\Delta_{pi}}{b}g_{1}(p)g_{2}(p)\\
\label{b.4Mesti}
&+q_{2}(i)g_{1}(i)g_{2}(i)\sum_{|p|\le{n}}\frac{\Delta_{pi}}{b}{\bf
E}\{G^{2}_{1}(p,p)\}g_{2}(p)+o(1), \quad \hbox{ as } \
n,b\rightarrow\infty.
\end{align}
If one considers (\ref{b.4Mesti}) with $i$ taken far enough from the
endpoints $-n$, $n$, then one can use relation (\ref{b.4supEGii})
and (\ref{b.4supq}) and replace $g_{1}$, $g_{2}$ and $q_{2}$  by the
constant values $w_{1}$ and $w_{2}$. This substitution leads to
simplified expressions with error terms that vanish as
$n,b\rightarrow \infty$. To finish the proof, we use relation
(\ref{b.2psicondit}) and Lemma 4.3 and replace the terms
$\sum_{p}\Delta_{pi}/b$ and $G^{2}_{1}$ of $M_{12}$ by the
corresponding expressions given by relations (\ref{b.2Delta}) and
(\ref{b.4supEG2ii}). This proves (\ref{b.4supUpsilon}). Lemma 4.2 is
proved. $\hfill \blacksquare$\vskip0,5cm

\subsubsection{Proof of Theorem 2.1}
Let us return to relation (\ref{b.4R12form5}). We introduce the
linear operator $W^{(g_{2},q_{2})}$ acting on vectors $e\in \C^{N}$
with components $e(i)$ as follows;
\begin{equation}
\{W^{(g_{2},q_{2})}(e)\}(i)=
v^{2}g_{2}(i)q_{2}(i)\sum_{|p|\le{n}}e(p)U(p,i).
\end{equation}
As a matter of fact, we can rewrite (\ref{b.4R12form5}) in the
following form:
\begin{equation}\label{b.4R12form6}
[I-W^{(g_{2},q_{2})}](R_{12})(i)=\frac{1}{Nb}f_{12}(i)+\Gamma(i),
\end{equation}
where $f_{12}$ and $\Gamma$ are given by (\ref{b.4f12}) and
(\ref{b.4Gamma}). It is easy to see that if $z\in\Lambda_{\eta}$,
then inequalities (\ref{b.3GijleImz}) and (\ref{b.4q2ileImz}) imply
that
\begin{equation*}
||W^{(g_{2},q_{2})}||_{1}\le{\frac{v^{2}}{(2v+1)^{2}}}<{\frac{1}{2}},
\end{equation*}
where
$||W^{(g_{2},q_{2})}||_{1}=\sup_{|V|_{1}=1}|W^{(g_{2},q_{2})}(V)|_{1}$
and $|V|_{1}=\sup_{i}|V(i)|$. Then (\ref{b.4R12form6}) can be
rewritten in the form
\begin{equation*}
R_{12}(i)=\frac{1}{Nb}\sum_{m=0}^{\infty}
\left([W^{(g_{2},q_{2})}]^{m}\vec{f}_{12}\right)(i)
+o\left(\frac{1}{Nb}\right).
\end{equation*}
Regarding the trace
\begin{equation*}
\frac{1}{N}\sum_{|i|\le{n}}R_{12}(i)=\frac{1}{N}\sum_{i\in
B_{L}}R_{12}(i)+\frac{2bL}{N}O\left(\frac{1}{Nb}\right) =
\frac{1}{N}\sum_{i\in B_{L}}R_{12}(i) + o\left(\frac{1}{Nb}\right)
\end{equation*}
and repeating the same arguments of the proof of (\ref{b.4supbF12U})
presented above, we can write that
\begin{equation*}
R_{12}(i)=\frac{1}{Nb}
\sum_{m=0}^{M}\sum_{|t|\le{n}}f_{12}(t)(v^{2}w^{2}_{2}U)^{m}(t,i)
+\frac{1}{Nb} \Delta^{(2)}(i)
\end{equation*}
with $\sup_{i\in B_{L}}|\Delta^{(2)}(i)|=o(1)$. Finally, observing
that $f_{12}(t)$ asymptotically does not depend on $t$ (see Lemma
4.2), we arrive with the help of (\ref{b.4supUj-1}), at the
expression (\ref{b.2Tz1z2}). Theorem 2.1 is proved. $\hfill
\blacksquare$

\section{Proof of auxiliary statement}
The main goal of this section is to prove Lemmas 3.1, 4.1 and 4.3.

\subsection{Proof of Lemma 3.1}
\subsubsection{Estimate of the term $Y_{1}$ (\ref{b.3R12form2})}
Here we have to use the resolvent identity (\ref{b.2h-zI-1}) and the
cumulants expansion formula (\ref{b.2EXjFX1Xm}) twice. However, the
computations are based on the same inequalities as those of the
proofs of Lemma 3.2. Regarding $Y_{1}(i)=\zeta_{2}b^{-2}\sum_{p}{\bf
E}\{g^{0}_{1}G_{2}(i,i)^{2}G_{2}(p,p)^{2}\}\Delta_{pi}$, we apply to
$G_{2}(i,i)$ the resolvent identity (\ref{b.2h-zI-1}). Then we get
relation
\begin{align*}
Y_{1}(i)=&{\zeta^{2}_{2}}{b^{2}}\sum_{|p|\le{n}}{\bf E}\{g_{1}^{0}
G_{2}(i,i)G_{2}(p,p)^{2}\}\Delta_{pi}\\
&-\frac{\zeta^{2}_{2}}{b^{2}}\sum_{|s|,|p|\le{n}}{\bf
E}\{g_{1}^{0}G_{2}(i,i)G_{2}(i,s)G_{2}(p,p)^{2}H(s,i)\}\Delta_{pi}.
\end{align*}
Applying the formula (\ref{b.2EXjFX1Xm}) with $q=3$ to ${\bf
E}\{g_{1}^{0}G_{2}(i,i)G_{2}(i,s)G_{2}(p,p)^{2}H(s,i)\}$, we obtain
that
\begin{align}
\nonumber Y_{1}(i)=&\frac{\zeta^{2}_{2}}{b^{2}}\sum_{|p|\le{n}}{\bf
E}(g_{1}^{0}G_{2}(i,i)G_{2}(p,p)^{2})\Delta_{pi}\\
\label{b.5Y1form1}
&+\frac{\zeta^{2}_{2}v^{2}}{b^{2}}\sum_{|p|\le{n}}{\bf
E}\{g_{1}^{0}G_{2}(i,i)^{2}G_{2}(p,p)^{2}U_{G_{2}}(i)\}\Delta_{pi}+
\sum_{r=1}^{3}Q_{r}(i)
\end{align}
with
\begin{align*}
Q_{1}(i)=&\frac{2\zeta^{2}_{2}v^{2}}{Nb^{2}}\sum_{|s|,|p|\le{n}}{\bf
E}\{G_{1}^{2}(i,s)G_{2}(i,s)G_{2}(i,i)G_{2}(p,p)^{2}\}U(s,i)\Delta_{pi}\\
&+\frac{3\zeta^{2}_{2}v^{2}}{b^{2}}\sum_{|s|,|p|\le{n}}{\bf
E}\{g_{1}^{0}G_{2}(i,i)G_{2}(p,p)^{2}G_{2}(i,s)^{2}\}U(s,i)\Delta_{pi}\\
&+\frac{4\zeta^{2}_{2}v^{2}}{b^{2}}\sum_{|s|,|p|\le{n}}{\bf
E}\{g_{1}^{0}G_{2}(i,i)G_{2}(p,p)G_{2}(i,s)
G_{2}(i,p)G_{2}(p,s)\}U(s,i)\Delta_{pi},\\
Q_{2}(i)=&-\frac{\zeta^{2}_{2}}{b^{2}}\sum_{|s|,|p|\le{n}}\frac{K_{4}}{6}
{\bf
E}\left\{D_{si}^{3}(g^{0}_{1}G_{2}(i,i)G_{2}(i,s)G_{2}(p,p)^{2})
\right\}\Delta_{pi}
\end{align*}
and
\begin{align*}
Q_{3}(i)=&-\frac{\zeta^{2}_{2}}{b^{2}4!}\sum_{|s|,|p|\le{n}}{\bf
E}\left\{H(s,i)^{5}[D_{si}^{4}(g^{0}_{1}G_{2}(i,i)G_{2}(i,s)
G_{2}(p,p)^{2})]^{(0)}\right\}\Delta_{pi}\\
&+\frac{\zeta^{2}_{2}}{b^{2}3!}\sum_{|s|,|p|\le{n}}K_{2}{\bf
E}\left\{H(s,i)^{3}[D_{si}^{4}(g^{0}_{1}G_{2}(i,i)G_{2}(i,s)
G_{2}(p,p)^{2})]^{(1)}\right\}\Delta_{pi}\\
&+\frac{\zeta^{2}_{2}}{b^{2}3!}\sum_{|s|,|p|\le{n}}K_{4}{\bf
E}\left\{H(s,i)[D_{si}^{4}(g^{0}_{1}G_{2}(i,i)G_{2}(i,s)
G_{2}(p,p)^{2})]^{(2)}\right\}\Delta_{pi},
\end{align*}
where $K_{r}$, $r=2,4$ are the cumulants of $H(s,i)$ as in
(\ref{b.3K2K4}). Applying to the second term of the RHS of
(\ref{b.5Y1form1}) identity (\ref{b.3Efg}) and using the definition
of $q_{2}(i)$ (\ref{b.4q2i}), we obtain that
\begin{align}
\nonumber
Y_{1}(i)=&\frac{\zeta_{2}q_{2}(i)}{b^{2}}\sum_{|p|\le{n}}{\bf
E}\{g_{1}^{0}G_{2}(i,i)G_{2}(p,p)^{2}\}\Delta_{pi}\\
\nonumber &+\frac{\zeta_{2}v^{2}q_{2}(i)}{b^{2}}\sum_{|p|\le{n}}{\bf
E}\{g_{1}^{0}G_{2}(i,i)^{2}G_{2}(p,p)^{2}U^{0}_{G_{2}}(i)\}\Delta_{pi}\\
\label{b.5Y1form2}
&+\frac{q_{2}(i)}{\zeta_{2}}\sum_{r=1}^{3}Q_{r}(i).
\end{align}
Regarding the first term of the RHS of this equality, we apply the
resolvent identity (\ref{b.2h-zI-1}) to $G_{2}(i,i)$. Repeating the
usual computations based on the formula (\ref{b.2EXjFX1Xm}) (with
$q=3$) and relation (\ref{b.2partialG}), we obtain that
\begin{equation*}
\frac{\zeta_{2}q_{2}(i)}{b^{2}}\sum_{|p|\le{n}}{\bf
E}\{g_{1}^{0}G_{2}(i,i)G_{2}(p,p)^{2}\}\Delta_{pi}
=\frac{\zeta_{2}}{b^{2}}q^{2}_{2}(i)\sum_{|p|\le{n}}{\bf
E}\{g_{1}^{0}G_{2}(p,p)^{2}\}\Delta_{pi}
\end{equation*}
\begin{equation}\label{b.5Y1form3}
+\frac{\zeta_{2}v^{2}}{b^{2}}q^{2}_{2}(i)\sum_{|p|\le{n}}{\bf
E}\{g_{1}^{0}G_{2}(i,i)G_{2}(p,p)^{2}U^{0}_{G_{2}}(i)\}\Delta_{pi}
+\frac{q_{2}(i)}{\zeta_{2}}\sum_{r=1}^{3}\breve{Q}_{r}(i)
\end{equation}
with
\begin{align*}
\breve{Q}_{1}(i)=
&\frac{2\zeta^{2}_{2}v^{2}q_{2}(i)}{Nb^{2}}\sum_{|s|,|p|\le{n}} {\bf
E}\{G_{1}^{2}(i,s)G_{2}(i,s)G_{2}(p,p)^{2}\}U(s,i)\Delta_{pi}\\
&+\frac{\zeta^{2}_{2}v^{2}q_{2}(i)}{b^{2}}\sum_{|s|,|p|\le{n}}{\bf
E}\{g_{1}^{0}G_{2}(p,p)^{2}G_{2}(i,s)^{2}\}U(s,i)\Delta_{pi}\\
&+\frac{4\zeta^{2}_{2}v^{2}q_{2}(i)}{b^{2}}\sum_{|s|,|p|\le{n}}{\bf
E}\{g_{1}^{0}G_{2}(p,p)G_{2}(i,s)
G_{2}(i,p)G_{2}(p,s)\}U(s,i)\Delta_{pi},\\
\breve{Q}_{2}(i)=&-\frac{\zeta^{2}_{2}q_{2}(i)}{b^{2}}\sum_{|s|,|p|\le{n}}\frac{K_{4}}{6}
{\bf E}\left\{D_{si}^{3}(g^{0}_{1}G_{2}(i,s)G_{2}(p,p)^{2})
\right\}\Delta_{pi}
\end{align*}
and
\begin{align*}
\breve{Q}_{3}(i)=
&-\frac{\zeta^{2}_{2}}{b^{2}4!}\sum_{|s|,|p|\le{n}}{\bf
E}\left\{H(s,i)^{5}[D_{si}^{4}(g^{0}_{1}G_{2}(i,s)
G_{2}(p,p)^{2})]^{(0)}\right\}\Delta_{pi}\\
&+\frac{\zeta^{2}_{2}}{b^{2}3!}\sum_{|s|,|p|\le{n}}K_{2}{\bf
E}\left\{H(s,i)^{3}[D_{si}^{4}(g^{0}_{1}G_{2}(i,s)
G_{2}(p,p)^{2})]^{(1)}\right\}\Delta_{pi}\\
&+\frac{\zeta^{2}_{2}}{b^{2}3!}\sum_{|s|,|p|\le{n}}K_{4}{\bf
E}\left\{H(s,i)[D_{si}^{4}(g^{0}_{1}G_{2}(i,s)
G_{2}(p,p)^{2})]^{(2)}\right\}\Delta_{pi},
\end{align*}
where $K_{r}$, $r=2,4$ are the cumulants of $H(s,i)$ as in
(\ref{b.3K2K4}).

Substituting (\ref{b.5Y1form3}) into (\ref{b.5Y1form2}), we obtain
that
\begin{align}
\nonumber
Y_{1}(i)=&\frac{\zeta_{2}}{b^{2}}q^{2}_{2}(i)\sum_{|p|\le{n}}{\bf
E}\{g_{1}^{0}G_{2}(p,p)^{2}\}\Delta_{pi}\\
\nonumber
&+\frac{\zeta_{2}v^{2}}{b^{2}}q^{2}_{2}(i)\sum_{|p|\le{n}}{\bf
E}\{g_{1}^{0}G_{2}(i,i)G_{2}(p,p)^{2}U^{0}_{G_{2}}(i)\}\Delta_{pi}\\
\nonumber &+\frac{\zeta_{2}v^{2}q_{2}(i)}{b^{2}}\sum_{|p|\le{n}}{\bf
E}\{g_{1}^{0}G_{2}(i,i)^{2}G_{2}(p,p)^{2}U^{0}_{G_{2}}(i)\}\Delta_{pi}\\
\label{b.5Y1form4}
&+\frac{q_{2}(i)}{\zeta_{2}}\sum_{r=1}^{3}Q_{r}(i)
+\frac{q_{2}(i)}{\zeta_{2}}\sum_{r=1}^{3}\breve{Q}_{r}(i).
\end{align}
Now let us estimate each term of the RHS of this equality. If one
assumes for a while that
\begin{equation}\label{b.5Y1estim1}
\sup_{|p|\le{n}}|{\bf E}\{g_{1}^{0}G_{2}(p,p)^{2}\}
|=O\left(N^{-1}b^{-1}+b^{-1}[{\bf Var}\{g_{1}\}]^{1/2}\right)
\end{equation}
holds for enough $n$ and $b$ satisfying (\ref{b.2bnalpha}). Then
this estimate and relations (\ref{b.4q2ileImz}), (\ref{b.4K4estim})
and (\ref{b.4supEUG02}) imply that the fist, the second and the
third terms of the RHS of (\ref{b.5Y1form4}) are of the order
indicated in the RHS of (\ref{b.3Y1Y2lem}).

Inequality (\ref{b.3GijleImz}), (\ref{b.3sumGij2leImz2}),
(\ref{b.3sumG1ipG2ip}) (with $m=1$ and $m=2$) and
(\ref{b.4q2ileImz}) imply that the term
$q_{2}(i)\zeta^{-1}_{2}[Q_{1}(i)+\breve{Q}_{1}(i)]$ is of the order
indicated in the RHS of (\ref{b.3Y1Y2lem}). Using similar arguments
as those of the proof of (\ref{b.3epsilonlem}) (see
(\ref{b.3epsilonestim1})-(\ref{b.3epsilonestim3})) and the following
estimates (cf. (\ref{b.3Vargnbnu})-(\ref{b.3D6pig10G2ip}))
\begin{equation*}
D_{si}^{r}(g^{0}_{1}G_{2}(i,i)G_{2}(i,s)G_{2}(p,p)^{2})=O\left(N^{-1}+|g_{1}^{0}|\right),
\quad r=3,4
\end{equation*}
and
\begin{equation*}
{\bf Var}\{[g_{n,b}(z_{l})]^{(\nu)}\}=O\left({\bf
Var}\{g_{n,b}(z_{l})\}+b^{-1}N^{-2}\right), \quad \nu=0,1,2,
\end{equation*}
we obtain that the terms $Q_{r}$, $r=2,3$ are of the order indicated
in the RHS of (\ref{b.3Y1Y2lem}). We conclude that the terms
$\breve{Q}_{r}$, $r=2,3$ and $\sup_{i}|Y_{1}(i)|$ are of the order
indicated in the RHS of (\ref{b.3Y1Y2lem}). \vskip0,5cm

Now let us prove (\ref{b.5Y1estim1}). Let us apply the resolvent
identity (\ref{b.2h-zI-1}) to $G_{2}(p,p)$. Repeating the usual
computations based on the formula (\ref{b.2EXjFX1Xm}) (with $q=3$)
and relation (\ref{b.2partialG}), we obtain that
\begin{align}
\nonumber {\bf E}\{g_{1}^{0}G_{2}(p,p)^{2}\}= &q_{2}(p){\bf
E}\{g_{1}^{0}G_{2}(p,p)\}+q_{2}(p){\bf
E}\{g_{1}^{0}G_{2}(p,p)^{2}U^{0}_{G_{2}}(p)\}\\
\nonumber &+3q_{2}(p)v^{2}\sum_{|s|\le{n}}{\bf
E}\{g_{1}^{0}G_{2}(p,s)^{2}G_{2}(p,p)\}U(s,p)\\
\nonumber &+\frac{2v^{2}}{N}q_{2}(p)\sum_{|s|\le{n}}{\bf
E}\{G^{2}_{1}(s,p)G_{2}(s,p)G_{2}(p,p)\}U(s,p) \\
\label{b.5Y1estim2} &-\frac{q_{2}(p)}{6}\sum_{|s|\le{n}}K_{4}{\bf
E}\left\{D_{sp}^{3}(g^{0}_{1}G_{2}(p,p)G_{2}(p,s))\right\}-q_{2}(p)
\tilde{Q}(i)
\end{align}
with
\begin{align*}
\tilde{Q}(i)=&-\frac{1}{4!}\sum_{|s|\le{n}}{\bf
E}\left\{H(s,p)^{5}[D_{sp}^{4}(g^{0}_{1}G_{2}(p,p)
G_{2}(p,s))]^{(0)}\right\}\\
&+\frac{1}{3!}\sum_{|s|\le{n}}K_{2}{\bf
E}\left\{H(s,p)^{3}[D_{sp}^{4}(g^{0}_{1}G_{2}(p,p)
G_{2}(p,s))]^{(1)}\right\}\\
&+\frac{1}{3!}\sum_{|s|\le{n}}K_{4}{\bf
E}\left\{H(s,p)[D_{sp}^{4}(g^{0}_{1}G_{2}(p,p)
G_{2}(p,s))]^{(2)}\right\},
\end{align*}
where $K_{r}$, $r=2,4$ are the cumulants of $H(s,p)$ as in
(\ref{b.3K2K4}).

Let us estimate each term of the RHS of (\ref{b.5Y1estim2}). It is
easy to show that the estimate of the first term of the RHS of
(\ref{b.5Y1estim2}) follows from the following statement, proved in
the previous work \cite{A}. \vskip0,2cm
\begin{lem}
(see \cite{A}) If $z\in\Lambda_{\eta}$, then under conditions of
Theorem 2.1, the estimate
\begin{equation}\label{b.5supEg10G2pp}
\sup_{|p|\le{n}}|{\bf
E}\{g_{1}^{0}G_{2}(p,p)\}|=O\left(b^{-1}n^{-1}+b^{-1}[{\bf
Var}\{g_{1}\}]^{1/2}\right)
\end{equation}
holds in the limit $n,b\rightarrow\infty$ (\ref{b.2bnalpha}).
\end{lem}
\vskip0,2cm

Then (\ref{b.5Y1estim1}) follows from this Lemma.and the estimate
(\ref{b.4supEUG02}) and the similar arguments used in the estimates
of the terms $Q_{r}$, $r=1,2,3$ in (\ref{b.5Y1form4}). Estimate
(\ref{b.5Y1estim1}) is proved. \vskip0,2cm

\subsubsection{Estimate of $Y_{2}$ (\ref{b.3R12form2})}
We rewrite $Y_{2}$ in the form $Y_{2}(i)=\zeta_{2}v^{2}\sum_{s}{\bf
E}\{M(i,s)\}U(s,i)$, where we denoted
\begin{equation*}
{\bf E}\{M(i,s)\}={\bf E}\{g^{0}_{1}G_{2}(i,s)^{2}\}.
\end{equation*}
To proceed with estimate of $Y_{2}$, we use the resolvent identity
(\ref{b.2h-zI-1}) and the cumulants expansion formula
(\ref{b.2EXjFX1Xm}) twice. However, the computations are based on
the results of Lemma 4.5, 4.6 and 5.1. Therefore we just indicate
the main lines of the proof and do not go into the details. Applying
to $G_{2}(i,s)$ the resolvent identity (\ref{b.2h-zI-1}), we get
equality
\begin{equation}\label{b.5Y2form1}
{\bf E}M(i,s)=\zeta_{2}\delta_{is}{\bf E}\{g^{0}_{1}G_{2}(i,i)\}
-\zeta_{2}\sum_{|t|\le{n}}{\bf
E}\{g^{0}_{1}G_{2}(i,s)G_{2}(i,t)H(t,s)\}.
\end{equation}
Regarding the first term of the RHS of this equality and using
relation (\ref{b.5supEg10G2pp}), it is easy to see that the term
\begin{equation*}
\sum_{|s|\le{n}}\zeta_{2}\delta_{is}{\bf
E}\{g^{0}_{1}G_{2}(i,i)\}U(s,i)=\zeta_{2}\frac{\psi(0)}{b}{\bf
E}\{g^{0}_{1}G_{2}(i,i)\}
\end{equation*}
is the value of order indicated in (\ref{b.3Y1Y2lem}). Let us
consider the second term of (\ref{b.5Y2form1}). Applying formula
(\ref{b.2EXjFX1Xm}) with $q=5$ to ${\bf
E}\{g^{0}_{1}G_{2}(i,s)G_{2}(i,t)H(t,s)\}$ and taking account
relations (\ref{b.2partialG}) and (\ref{b.3Efg}), we obtain that
\begin{equation}\label{b.5Y2form2}
-\zeta_{2}\sum_{|t|\le{n}}{\bf
E}\{g^{0}_{1}G_{2}(i,s)G_{2}(i,t)H(t,s)\}
=\sum_{l=1}^{7}\Theta_{l}(i,s),
\end{equation}
where
\begin{align*}
\Theta_{1}(i,s)=&v^{2}\zeta_{2}{\bf
E}\{g^{0}_{1}G_{2}(i,s)^{2}\}{\bf E}U_{G_{2}}(s),\\
\Theta_{2}(i,s)=&v^{2}\zeta_{2}{\bf
E}\{g^{0}_{1}G_{2}(i,s)^{2}U^{0}_{G_{2}}(s)\},\\
\Theta_{3}(i,s)=&\frac{2v^{2}\zeta_{2}}{N}\sum_{|t|\le{n}}{\bf
E}\{G^{2}_{1}(s,t)U(t,s)G_{2}(i,s)G_{2}(i,t)\},\\
\Theta_{4}(i,s)=&v^{2}\zeta_{2}\sum_{|t|\le{n}}{\bf
E}\{g^{0}_{1}G_{2}(i,t)^{2}G_{2}(s,s)\}U(t,s),\\
\Theta_{5}(i,s)=&2v^{2}\zeta_{2}\sum_{|t|\le{n}}{\bf
E}\{g^{0}_{1}G_{2}(i,s)G_{2}(t,s)G_{2}(i,t)\}U(t,s),\\
\Theta_{6}(i,s)=&-\zeta_{2}\sum_{|t|\le{n}}\frac{K_{4}\left(H(t,s)\right)}{6}
{\bf E}\{D^{3}_{ts}(g^{0}_{1}G_{2}(i,s)G_{2}(i,t))\}
\end{align*}
and
\begin{equation*}
\Theta_{7}(i,s)=-\zeta_{2}\sum_{|t|\le{n}}\frac{K_{6}\left(H(t,s)\right)}{5!}{\bf
E}\{D^{5}_{ts}(g^{0}_{1}G_{2}(i,s)G_{2}(i,t))\}+\tilde{\Theta}_{7}(i,s)
\end{equation*}
with
\begin{align*}
\tilde{\Theta}_{7}(i,s)=&-\frac{\zeta_{2}}{6!}\sum_{|t|\le{n}}{\bf
E}\left\{H(t,s)^{7}[D_{ts}^{6}(g^{0}_{1}G_{2}(i,s)G_{2}(i,t))]^{(0)}
\right\}\\
&+\frac{\zeta_{2}}{5!}\sum_{|t|\le{n}}K_{2}\left(H(t,s)\right){\bf
E}\left\{H(t,s)^{5}[D_{ts}^{6}(g^{0}_{1}G_{2}(i,s)G_{2}(i,t))]^{(1)}
\right\}\\
&+\frac{\zeta_{2}}{(3!)^{2}}\sum_{|t|\le{n}}K_{4}\left(H(t,s)\right){\bf
E}\left\{H(t,s)^{3}[D_{ts}^{6}(g^{0}_{1}G_{2}(i,s)G_{2}(i,t))]^{(2)}
\right\}\\
&+\frac{\zeta_{2}}{5!}\sum_{|t|\le{n}}K_{6}\left(H(t,s)\right){\bf
E}\left\{H(t,s)[D_{ts}^{6}(g^{0}_{1}G_{2}(i,s)G_{2}(i,t))]^{(3)}
\right\},
\end{align*}
where $K_{r}\left(H(t,s)\right)$, $r=2,4,6$ are the cumulants of
$H(t,s)$ as in (\ref{b.3K2K4})-(\ref{b.3K6}).

The term $\Theta_{1}$ is of the form $v^{2}\zeta_{2}{\bf
E}\{M(i,s)\}{\bf E}U_{G_{2}}(s)$ and can be put to the left hand
side of (\ref{b.5Y2form1}). The terms $\Theta_{2}$ and $\Theta_{3}$
are of the order indicated in the RHS of (\ref{b.3Y1Y2lem}). This
can be shown with the help of the estimate (\ref{b.4supEUG02}) and
inequality (eg. \cite{A})
\begin{align}
\nonumber &\left|\sum_{|s|,|t|\le{n}}G^{2}_{1}(s,t)G_{2}(i,s)G_{2}(i,t)\right| \\
\label{b.5sumstG1G2}
&\le{||G_{1}^{2}||\left(\sum_{|s|\le{n}}|G_{2}(i,s)|^{2}\right)^{1/2}
\left(\sum_{|t|\le{n}}|G_{2}(i,t)|^{2}\right)^{1/2}}
\le{\frac{1}{\eta^{4}}}.
\end{align}
Regarding $\Theta_{4}$, we apply the resolvent identity
(\ref{b.2h-zI-1}) to the factor $G_{2}(s,s)$. Repeating the usual
computations based on the formula (\ref{b.2EXjFX1Xm}) with $q=5$ and
taking into account relations (\ref{b.2partialG}) and
(\ref{b.3Efg}), we obtain that
\begin{equation}\label{b.5Theta4}
\Theta_{4}(i,s)=v^{2}\zeta^{2}_{2}\sum_{|t|\le{n}}{\bf
E}\{M(i,t)\}U(t,s) +v^{2}\zeta_{2}\Theta_{4}(i,s){\bf
E}U_{G_{2}}(s)+\sum_{l=1}^{8}\Omega_{l}(i,s),
\end{equation}
where
\begin{align*}
\Omega_{1}(i,s)=&v^{4}\zeta^{2}_{2}\sum_{|t|\le{n}}{\bf
E}\{g^{0}_{1}G_{2}(i,t)^{2}G_{2}(s,s)U^{0}_{G_{2}}(s)\}U(t,s),\\
\Omega_{2}(i,s)=&v^{4}\zeta^{2}_{2}\sum_{|t|,|p|\le{n}}{\bf
E}\{g_{1}^{0}G_{2}(i,t)^{2}G_{2}(s,p)^{2}\}U(p,s)U(t,s),\\
\Omega_{3}(i,s)=&\frac{2v^{4}\zeta^{2}_{2}}{N}\sum_{|t|,|p|\le{n}}{\bf
E}\{G^{2}_{1}(p,s)G_{2}(i,t)^{2}G_{2}(s,p)\}U(p,s)U(t,s),\\
\Omega_{4}(i,s)=&2v^{4}\zeta^{2}_{2}\sum_{|t|,|p|\le{n}}{\bf
E}\{g^{0}_{1}G_{2}(i,t)G_{2}(i,s)G_{2}(p,t)G_{2}(s,p)\}U(p,s)U(t,s),\\
\Omega_{5}(i,s)=&2v^{4}\zeta^{2}_{2}\sum_{|t|,|p|\le{n}}{\bf
E}\{g^{0}_{1}G_{2}(i,t)G_{2}(i,p)G_{2}(s,t)G_{2}(s,p)\}U(p,s)U(t,s),\\
\Omega_{6}(i,s)=&-\zeta^{2}_{2}v^{2}\sum_{|t|,|p|\le{n}}\frac{K_{4}\left(H(p,s)\right)}{6}
{\bf E}\{D^{3}_{ps}(g^{0}_{1}G_{2}(i,t)^{2}G_{2}(s,p))\}U(t,s),\\
\Omega_{7}(i,s)=&-\zeta^{2}_{2}v^{2}\sum_{|t|,|p|\le{n}}\frac{K_{6}\left(H(p,s)\right)}{5!}{\bf
E}\{D^{5}_{ps}(g^{0}_{1}G_{2}(i,t)^{2}G_{2}(s,p))\}U(t,s)
\end{align*}
and
\begin{align*}
\Omega_{8}(&i,s)\\
=&-\frac{\zeta^{2}_{2}v^{2}}{6!}\sum_{|t|,|p|\le{n}}{\bf
E}\left\{H(p,s)^{7}[D_{ps}^{6}(g^{0}_{1}G_{2}(i,t)^{2}G_{2}(s,p))]^{(0)}
\right\}U(t,s)\\
&+\frac{\zeta^{2}_{2}v^{2}}{5!}\sum_{|t|,|p|\le{n}}K_{2}\left(H(p,s)\right){\bf
E}\left\{H(p,s)^{5}[D_{ts}^{6}(g^{0}_{1}G_{2}(i,t)^{2}G_{2}(s,p))]^{(1)}
\right\}U(t,s)\\
&+\frac{\zeta^{2}_{2}v^{2}}{(3!)^{2}}\sum_{|t|,|p|\le{n}}K_{4}\left(H(p,s)\right){\bf
E}\left\{H(p,s)^{3}[D_{ts}^{6}(g^{0}_{1}G_{2}(i,t)^{2}G_{2}(s,p))]^{(2)}
\right\}U(t,s)\\
&+\frac{\zeta^{2}_{2}v^{2}}{5!}\sum_{|t|,|p|\le{n}}K_{6}\left(H(p,s)\right){\bf
E}\left\{H(p,s)[D_{ts}^{6}(g^{0}_{1}G_{2}(i,t)^{2}G_{2}(s,p))]^{(3)}
\right\}U(t,s)
\end{align*}
with $K_{r}\left(H(p,s)\right)$, $r=2,4,6$ are the cumulants of
$H(p,s)$ as in (\ref{b.3K2K4})-(\ref{b.3K6}).

The terms $\Omega_{l}$, $l=1,\ldots,5$ are of the order indicated in
the RHS of (\ref{b.3Y1Y2lem}). This can be shown with the help of
the estimate (\ref{b.4supEUG02}) and the inequalities
(\ref{b.3GijleImz}), (\ref{b.3sumGij2leImz2}),
(\ref{b.3sumG1ipG2ip}) and (\ref{b.5sumstG1G2}). The term
$\Omega_{6}$ contains $272$ terms that are of the order indicated in
the RHS of (\ref{b.3Y1Y2lem}). This can be checked by direct
computations with the use of (\ref{b.3sumG1ipG2ip}) and
(\ref{b.5sumstG1G2}). Using similar argument as those of the proofs
of (\ref{b.3Y3Y6lem}) and (\ref{b.3epsilonlem}) (see
(\ref{b.3epsilonestim1})-(\ref{b.3epsilonestim3})), and the
following estimate (cf. (\ref{b.3ED5pig10G2ij}))
\begin{equation}\label{b.5ED5}
{\bf
E}|D^{5}_{pi}\{g^{0}_{1}G_{2}(i,t)^{2}G_{2}(s,p)\}|=O\left(N^{-1}+[{\bf
Var}\{g_{1}\}]^{1/2}\right), \ \hbox{ as } \quad
n,p\rightarrow\infty,
\end{equation}
we conclude that the terms $\Omega_{7}$ and $\Omega_{8}$ are of the
order indicated in the RHS of (\ref{b.3Y1Y2lem}). Then, the relation
(\ref{b.5Theta4}) is of the form that leads to the estimates needed
for $\sum_{s}{\bf E}\{M(i,s)\}U(s,i)$.

Regarding $\Theta_{5}(i,s)$, we apply the resolvent identity
(\ref{b.2h-zI-1}) to the factor $G_{2}(t,s)$. Repeating the usual
computations based on the formula (\ref{b.2EXjFX1Xm}) with $q=5$ and
taking into account relations (\ref{b.2partialG}) and
(\ref{b.3Efg}), we obtain that
\begin{equation}\label{b.5Theta5}
\Theta_{5}(i,s)=2v^{2}\zeta^{2}_{2}\frac{\psi(0)}{b}{\bf E}M(i,s)
+v^{2}\zeta_{2}\Theta_{5}(i,s){\bf
E}U_{G_{2}}(s)+\sum_{l=1}^{9}\Omega^{'}_{l}(i,s),
\end{equation}
where
\begin{align*}
\Omega^{'}_{1}(i,s)=&2v^{4}\zeta^{2}_{2}\sum_{|t|\le{n}}{\bf
E}\{g^{0}_{1}G_{2}(i,s)G_{2}(i,t)G_{2}(t,s)U^{0}_{G_{2}}(s)\}U(t,s),\\
\Omega^{'}_{2}(i,s)=&2v^{4}\zeta^{2}_{2}\sum_{|t|,|p|\le{n}}{\bf
E}\{g_{1}^{0}G_{2}(i,p)G_{2}(s,s)G_{2}(i,t)G_{2}(t,p)\}U(p,s)U(t,s),\\
\Omega^{'}_{3}(i,s)=&\frac{4v^{4}\zeta^{2}_{2}}{N}\sum_{|t|,|p|\le{n}}{\bf
E}\{G^{2}_{1}(p,s)G_{2}(i,s)G_{2}(i,t)G_{2}(t,p)\}U(p,s)U(t,s),\\
\Omega^{'}_{4}(i,s)=&4v^{4}\zeta^{2}_{2}\sum_{|t|,|p|\le{n}}{\bf
E}\{g^{0}_{1}G_{2}(i,s)G_{2}(p,s)G_{2}(i,t)G_{2}(t,p)\}U(p,s)U(t,s),\\
\Omega^{'}_{5}(i,s)=&2v^{4}\zeta^{2}_{2}\sum_{|t|,|p|\le{n}}{\bf
E}\{g^{0}_{1}G_{2}(i,s)G_{2}(i,p)G_{2}(t,p)G_{2}(s,t)\}U(p,s)U(t,s),\\
\Omega^{'}_{6}(i,s)=&2v^{4}\zeta^{2}_{2}\sum_{|t|,|p|\le{n}}{\bf
E}\{g^{0}_{1}G_{2}(i,s)^{2}G_{2}(p,t)^{2}\}U(p,s)U(t,s),\\
\Omega^{'}_{7}(i,s)=&-2\zeta^{2}_{2}v^{2}\sum_{|t|,|p|\le{n}}\frac{K_{4}\left(H(p,s)\right)}{6}
{\bf E}\{D^{3}_{ps}(g^{0}_{1}G_{2}(i,s)G_{2}(i,t)G_{2}(t,p))\}U(t,s),\\
\Omega^{'}_{8}(i,s)=&-2\zeta^{2}_{2}v^{2}\sum_{|t|,|p|\le{n}}\frac{K_{6}\left(H(p,s)\right)}{5!}{\bf
E}\{D^{5}_{ps}(g^{0}_{1}G_{2}(i,s)G_{2}(i,t)G_{2}(t,p))\}U(t,s)
\end{align*}
and
\begin{align*}
\Omega^{'}&_{9}(i,s)\\
=&-\frac{2\zeta^{2}_{2}v^{2}}{6!}\sum_{|t|,|p|\le{n}}{\bf
E}\left\{H(p,s)^{7}[D_{ps}^{6}(g^{0}_{1}G_{2}(i,s)G_{2}(i,t)G_{2}(t,p))]^{(0)}
\right\}U(t,s)\\
&+\frac{2\zeta^{2}_{2}v^{2}}{5!}\sum_{|t|,|p|\le{n}}K_{2}\left(H(p,s)\right){\bf
E}\left\{H(p,s)^{5}[D_{ts}^{6}(g^{0}_{1}G_{2}(i,s)G_{2}(i,t)G_{2}(t,p))]^{(1)}
\right\}U(t,s)\\
&+\frac{2\zeta^{2}_{2}v^{2}}{(3!)^{2}}\sum_{|t|,|p|\le{n}}K_{4}\left(H(p,s)\right){\bf
E}\left\{H(p,s)^{3}[D_{ts}^{6}(g^{0}_{1}G_{2}(i,s)G_{2}(i,t)G_{2}(t,p))]^{(2)}
\right\}U(t,s)\\
&+\frac{2\zeta^{2}_{2}v^{2}}{5!}\sum_{|t|,|p|\le{n}}K_{6}\left(H(p,s)\right){\bf
E}\left\{H(p,s)[D_{ts}^{6}(g^{0}_{1}G_{2}(i,s)G_{2}(i,t)G_{2}(t,p))]^{(3)}
\right\}U(t,s)
\end{align*}
with $K_{r}\left(H(p,s)\right)$, $r=2,4,6$ are the cumulants of
$H(p,s)$ as in (\ref{b.3K2K4})-(\ref{b.3K6}).

The terms $\sum_{s}\Omega^{'}_{l}(i,s)U(s,i)$, $l=1,\ldots,6$ are of
the order indicated in the RHS of (\ref{b.3Y1Y2lem}). This can be
shown with the help of the estimate (\ref{b.4supEUG02}) and the
inequalities (\ref{b.3GijleImz}), (\ref{b.3sumGij2leImz2}),
(\ref{b.3sumG1ipG2ip}) and (\ref{b.5sumstG1G2}). The term
$\Omega^{'}_{7}$ contains $356$ terms that are of the order
indicated in the RHS of (\ref{b.3Y1Y2lem}). This can be checked by
direct computations with the use of (\ref{b.3sumG1ipG2ip}) and
(\ref{b.5sumstG1G2}). Using similar argument as those of the proofs
of (\ref{b.3Y3Y6lem}) and (\ref{b.3epsilonlem}) (see
(\ref{b.3epsilonestim1})-(\ref{b.3epsilonestim3})), and the
following estimate (cf. (\ref{b.3ED5pig10G2ij}))
\begin{equation}\label{b.5ED5}
{\bf
E}|D^{5}_{pi}\{g^{0}_{1}G_{2}(i,s)G_{2}(i,t)G_{2}(t,p)\}|=O\left(N^{-1}+[{\bf
Var}\{g_{1}\}]^{1/2}\right), \ \hbox{ as } \quad
n,p\rightarrow\infty,
\end{equation}
we conclude that the terms $\Omega^{'}_{8}$ and $\Omega^{'}_{9}$ are
of the order indicated in the RHS of (\ref{b.3Y1Y2lem}). Then, the
form of (\ref{b.5Theta5}) is also such that, being substituted into
(\ref{b.5Y2form2}) and then into (\ref{b.5Y2form1}), it leads to the
needed estimates.

The term $\Theta_{6}(i,s)$ contains $67$ terms. These terms can be
gathered into three groups. In each group, the terms are estimated
by the same values with the help of the same computations.

We give estimates for the typical cases. Using (\ref{b.3GijleImz}),
(\ref{b.3sumGij2leImz2}) and (\ref{b.3sumG1ipG2ip}) (with $m=1$), we
get for the terms of the first group:
\begin{align*}
&\left|\frac{\zeta_{2}}{N}\sum_{|t|\le{n}}K_{4}\left(H(t,s)\right){\bf
E}\{G_{1}^{2}(t,t)G_{1}(s,s)G_{1}(t,s)G_{2}(i,s)G_{2}(i,t)\}\cdot
\frac{1}{(1+\delta_{ts})^{3}}\right|\\
&\le{\frac{V_{4}+3v^{4}}{\eta^{5}Nb^{2}}\sum_{|t|\le{n}}{\bf
E}|G_{1}(t,s)G_{2}(i,t)|}\le{\frac{V_{4}+3v^{4}}{\eta^{7}Nb^{2}}}.
\end{align*}
For the terms of the second group, we obtain estimates
\begin{align*}
&\left|\zeta_{2}\sum_{|t|\le{n}}K_{4}\left(H(t,s)\right){\bf
E}\{g_{1}^{0}G_{2}(s,s)^{2}G_{2}(i,t)^{2}\}\cdot
\frac{1}{(1+\delta_{ts})^{3}}\right|\\
&\le{\frac{V_{4}+3v^{4}}{\eta^{3}b^{2}}{\bf
E}|g_{1}^{0}|\sum_{|t|\le{n}}|G_{2}(i,t)^{2}|}\le{\frac{[V_{4}+3v^{4}]\sqrt{{\bf
Var}\{g_{1}\}}}{\eta^{5}b^{2}}}.
\end{align*}
Finally, for the terms of the third group, we get inequalities
\begin{align*}
&\left|\sum_{|s|\le{n}}\frac{\zeta_{2}}{N}\sum_{|t|\le{n}}K_{4}\left(H(t,s)\right){\bf
E}\{G_{1}^{2}(s,s)G_{1}(t,t)G_{2}(t,t)G_{2}(i,s)^{2}\}U(s,i)\cdot
\frac{1}{(1+\delta_{ts})^{3}}\right|\\
&\le{\frac{V_{4}+3v^{4}}{\eta^{5}Nb}\sum_{|s|\le{n}}{\bf
E}|G_{2}(i,s)^{2}|\sum_{|t|\le{n}}U(t,s)U(s,i)}=O\left(\frac{1}{Nb^{2}}\right).
\end{align*}
Gathering all the estimates of $67$ terms, we obtain that
\begin{equation*}
\left|\sum_{|s|\le{n}}\Theta_{6}(i,s)U(s,i)\right|=O\left(\frac{1}{Nb^{2}}+\frac{\sqrt{{\bf
Var}\{g_{1}\}}}{b^{2}}\right).
\end{equation*}

Using similar argument as those of the proofs of (\ref{b.3Y3Y6lem})
and (\ref{b.3epsilonlem}) (see
(\ref{b.3epsilonestim1})-(\ref{b.3epsilonestim3})), we conclude that
$\Theta_{7}$ and $\sup_{i}|Y_{2}(i)|$ are of the order indicated in
(\ref{b.3Y1Y2lem}). Estimate (\ref{b.3Y1Y2lem}) is proved and so
Lemma 3.1 is proved.$\hfill \blacksquare$

\vskip0,5cm
\subsection{Proof of Lemma 4.1}
Let us consider the variable
\begin{equation*}
K(i,s)={\bf E}\{RG^{0}(i,i)\}={\bf E}\{R^{0}G(i,i)\},
\end{equation*}
where we denoted $R=g^{0}U^{0}_{G}(s)$. Applying to $G_{2}(i,i)$ the
resolvent identity (\ref{b.2h-zI-1}) and taking account formula
(\ref{b.2EXjFX1Xm}) with $q=3$ and relation (\ref{b.2partialG}), we
obtain that
\begin{equation}\label{b.5RG}
{\bf E}\{R^{0}G(i,i)\}=\zeta v^{2}{\bf E}\{R^{0}G(i,i)U_{G}(i)\} +
\sum_{a=1}^{5}l_{a}(i,s)
\end{equation}
with
\begin{align*}
l_{1}(i,s)=&\zeta v^{2}\sum_{|p|\le{n}}{\bf
E}\{R^{0}G(i,p)^{2}\}U(p,i),\\
l_{2}(i,s)=&2\zeta v^{2}\sum_{|p|,|t|\le{n}}{\bf
E}\{g^{0}G(t,p)G(t,i)G(i,p)\}U(t,s)U(p,i),\\
l_{3}(i,s)=&\frac{2\zeta v^{2}}{N}\sum_{|p|,|t|\le{n}}{\bf
E}\{G(p,t)G(i,t)U^{0}_{G}(s)G(i,p)\}U(p,i),\\
l_{4}(i,s)=&-\frac{\zeta}{6}\sum_{|p|\le{n}}K_{4}{\bf
E}\{D^{3}_{pi}(R^{0}G(i,p))\}
\end{align*}
and
\begin{align*}
l_{5}(i,s)=&-\frac{\zeta}{4!}\sum_{|p|\le{n}}{\bf
E}\left\{H(p,i)^{5}[D_{pi}^{4}(R^{0}G(i,p))]^{(0)}\right\}\\
&+\frac{\zeta}{3!}\sum_{|p|\le{n}}K_{2}{\bf
E}\left\{H(p,i)^{3}[D_{pi}^{4}(R^{0}G(i,p))]^{(1)}\right\}\\
&+\frac{\zeta}{3!}\sum_{|p|\le{n}}K_{4}{\bf
E}\left\{H(p,i)[D_{pi}^{4}(R^{0}G(i,p))]^{(2)}\right\},
\end{align*}
where $K_{r}$, $r=2,4$ are the cumulants of $H(p,i)$ as in
(\ref{b.3K2K4}). Let us use the identity
\begin{equation*}
{\bf E}R^{0}XY={\bf E}RX^{0}{\bf E}Y+{\bf E}RY^{0}{\bf E}X+{\bf
E}RX^{0}Y^{0}-{\bf E}R{\bf E}X^{0}Y^{0},
\end{equation*}
and rewrite (\ref{b.5RG}) in the form
\begin{equation}\label{b.5RGform1}
{\bf
E}\{R^{0}G(i,i)\}=K(i,s)=v^{2}q(i)g(i)\sum_{|t|\le{n}}K(t,s)U(t,i)+
\Pi(i,s)
\end{equation}
with
\begin{align}
\nonumber \Pi(i,s)=&v^{2}q(i)\left[{\bf
E}\{RU^{0}_{G}(i)G^{0}(i,i)\}-{\bf
E}\{g^{0}U^{0}_{G}(s)\}{\bf E}\{G^{0}(i,i)U^{0}_{G}(i)\}\right]\\
\label{b.5Pi} &+ \frac{q(i)}{\zeta}\sum_{a=1}^{5}l_{a}(i,s),
\end{align}
where $g(i)={\bf E}\{G(i,i)\}$ and $q$ is given by (\ref{b.4q2i}).
Now we rewrite (\ref{b.5RGform1}) in the form of a vector equality
\begin{equation*}
\vec{K}(.,s)=[I-W^{(q,g)}]^{-1}\vec{\Pi}(.,s),
\end{equation*}
where we denote by $W^{(q,g)}$ the linear operator acting on a
vector $e$ with components $e(i)$ as
\begin{equation*}
[W^{(q,g)}e](i)=v^{2}q(i)g(i)\sum_{|t|\le{n}}e(t)U(t,i)
\end{equation*}
and vectors $[\vec{\Pi}(.,s)](i)=\Pi(i,s)$. It is easy to see that
if $z\in \Lambda_{\eta}$, then $||W^{(q,g)}||\le{\frac{1}{2}}$.
Thus, to prove relation (\ref{b.4taulem}), it is sufficient to show
that
\begin{equation}\label{b.5Piestim}
\sup_{|i|,|s|\le{n}}|\Pi(i,s)|=O\left(\frac{1}{Nb^{2}}
+\frac{1}{b^{2}}\left({\bf Var}\{g\}\right)^{1/2}\right).
\end{equation}
Let us prove (\ref{b.5Piestim}). Taking into account inequality
(\ref{b.3GijleImz}), (\ref{b.3sumGij2leImz2}), (\ref{b.5sumstG1G2})
and estimate (\ref{b.4supEUG02}), we obtain that
\begin{equation}
|l_{a}(i,s)|\le{\frac{c}{b^{2}}\left({\bf Var}\{g\}\right)^{1/2}}
\quad \ a=1,2
\end{equation}
and
\begin{equation}
|l_{3}(i,s)|\le{\frac{c}{Nb^{2}}},
\end{equation}
where c is a constant. Using similar arguments as those of the proof
of (\ref{b.3epsilonlem}) (see
(\ref{b.3epsilonestim1})-(\ref{b.3epsilonestim3})) and the following
estimates (cf. (\ref{b.3Vargnbnu})-(\ref{b.3D6pig10G2ip}))
\begin{equation*}
D_{pi}^{r}(R^{0}G(i,p))=O\left(N^{-1}+|g_{1}^{0}|\right), \quad
r=3,4
\end{equation*}
and
\begin{equation*}
{\bf Var}\{[g_{n,b}(z)]^{(\nu)}\}=O\left({\bf
Var}\{g_{n,b}(z)\}+b^{-1}N^{-2}\right), \quad \nu=0,1,2,
\end{equation*}
we obtain that the terms $l_{a}$, $a=4,5$ are of the order indicated
in the RHS of (\ref{b.4taulem}). Finally, we derive inequality
\begin{align}
\nonumber |\Pi(i,s)|\le&{c\left({\bf
Var}\{g\}\right)^{1/2}\left(\left({\bf
E}|U^{0}_{G}(i)|^{4}\right)^{1/2}+\frac{1}{b^{2}}\left({\bf
E}|U^{0}_{G}(i)|^{2}\right)^{1/2}\right)}\\
\label{b.5Piestim2}
&+c\left(\frac{1}{Nb^{2}}+\frac{1}{b^{2}}\left({\bf
Var}\{g\}\right)^{1/2}\right),
\end{align}
where $c$ is a constant. Then (\ref{b.5Piestim}),
(\ref{b.5Piestim2}) and Lemma 4.1 follow from (\ref{b.4supEUG02})
and the following estimate. \vskip0,2cm
\begin{lem}
If $z\in\Lambda_{\eta}$, then under conditions of Theorem 2.1, the
estimate
\begin{equation}\label{b.4supEUG04}
\sup_{|s|\le{n}}{\bf E}\{|U^{0}_{G}(s;z)|^{4}\}=O(b^{-4})
\end{equation}
holds in the limit $n,b\rightarrow\infty$.
\end{lem}
\vskip0,2cm

{\it Proof of Lemma 5.2.} Let us consider variable
\begin{equation*}
{\bf
E}\{U^{0}_{G_{1}}(x_{1})U^{0}_{G_{2}}(x_{2})U^{0}_{G_{3}}(x_{3})U^{0}_
{G_{4}}(x_{4})\}= {\bf
E}[U^{0}_{G_{1}}(x_{1})U^{0}_{G_{2}}(x_{2})U^{0}_{G_{3}}
(x_{3})]^{0}U_{G_{4}}(x_{4}).
\end{equation*}
Set $T=U^{0}_{G_{1}}U^{0}_{G_{2}}U^{0}_{G_{3}}$ and
$M(x_{1},x_{2},x_{3},t)={\bf E}T^{0}G_{4}(t,t)$. We apply to
$G_{4}(t,t)$ the resolvent identity $(3.2)$ and obtain
\begin{equation*}
{\bf E}T^{0}G_{4}(t,t)=-\zeta_{4}\sum_{|s|\le{n}}{\bf
E}\{T^{0}G_{4}(t,s)H(s,t)\}.
\end{equation*}
Applying (\ref{b.2EXjFX1Xm}) to ${\bf E}\{T^{0}G_{4}(t,s)H(s,t)\}$
with $q=3$ and taking into account (\ref{b.2partialG}), we get
relation
\begin{align}
\nonumber &{\bf E}T^{0}G_{4}(t,t)\\
\nonumber &= \zeta_{4}v^{2}{\bf
E}\{T^{0}G_{4}(t,t)U_{G_{4}}(t)\}+\zeta_{4}v^{2}{\bf
E}\left\{T^{0}\sum_{|s|\le{n}}G_{4}(t,s)^{2}U(s,t)\right\}\\
\nonumber &+ 2\zeta_{4}v^{2}\sum_{(i,j,k)}{\bf
E}\left\{U^{0}_{G_{i}}(x_{i})U^{0}_{G_{j}}(x_{j})\sum_{|y|,|s|\le{n}}
G_{k}(y,s)G_{k}(t,y)G_{4}(t,s)U(y,x_{k})U(s,t)\right\}\\
\label{b.5W4form1} &+\zeta_{4}\Gamma_{1}(t)+\zeta_{4}\Gamma_{2}(t)
\end{align}
with
\begin{equation}\label{b.5daleth1}
\Gamma_{1}(t)=-\sum_{|s|\le{n}}\frac{K_{4}}{3!}{\bf
E}\left\{D^{3}_{st}(T^{0}G_{4}(t,s))\right\}
\end{equation}
and
\begin{align}
\nonumber \Gamma_{2}(t)=&-\frac{1}{4!}\sum_{|s|\le{n}}{\bf
E}\left\{H(s,t)^{5}[D^{4}_{st}\left(T^{0}G_{4}(t,s)\right)]^{(0)}
\right\}\\
\nonumber &+\sum_{|s|\le{n}}\frac{K_{2}}{3!}{\bf
E}\left\{H(s,t)^{3}[D^{4}_{st}\left(T^{0}G_{4}(t,s)\right)]^{(1)}
\right\} \\
\label{b.5daleth2} &+\sum_{|s|\le{n}}\frac{K_{4}}{3!}{\bf
E}\left\{H(s,t)[D^{4}_{st}\left(T^{0}G_{4}(t,s)\right)]^{(2)}
\right\},
\end{align}
where $K_{r}$, $r=2,4$ are the cumulants of $H(s,t)$ as in
(\ref{b.3K2K4}). In (\ref{b.5W4form1}), we introduce the notation
\begin{equation*}
\sum_{(i,j,k)}\xi(x_{i},x_{j},x_{k}) = \xi(x_{1},x_{2},x_{3})
+\xi(x_{1},x_{3},x_{2}) + \xi(x_{2},x_{3},x_{1}).
\end{equation*}
Applying to the first term of the RHS of (\ref{b.5W4form1}) relation
(\ref{b.3Efg}) and using $q_{4}(t)$ (\ref{b.4q2i}), we obtain that
\begin{align*}
&{\bf E}T^{0}G_{4}(t,t)\\
&= q_{4}(t) v^{2}{\bf E}\{T^{0}G_{4}(t,t)U^{0}_{G_{4}}(t)\}
+q_{4}(t)v^{2}{\bf
E}\left\{T^{0}\sum_{|s|\le{n}}G_{4}(t,s)^{2}U(s,t)\right\}\\
&+2q_{4}(t)v^{2}\sum_{(i,j,k)}{\bf
E}\left\{U^{0}_{G_{i}}(x_{i})U^{0}_{G_{j}}(x_{j})\sum_{|y|,|s|\le{n}}
G_{k}(y,s)G_{k}(t,y)G_{4}(t,s)U(y,x_{k})U(s,t)\right\}\\
&+q_{4}(t)\left(\Gamma_{1}(t)+\Gamma_{2}(t)\right).
\end{align*}
Now gathering relation given by (\ref{b.2psicondit}),
(\ref{b.3GijleImz}), (\ref{b.3sumGij2leImz2}), (\ref{b.5sumstG1G2}),
(\ref{b.4q2ileImz}) and
\begin{equation*}
\sup_{|t|\le{n}}{\bf E}|T^{0}U^{0}_{G_{4}}(t)|\le{{\bf
E}|T|\sup_{|t|\le{n}}{\bf E}|U^{0}_{G_{4}}(t)|}
+\sup_{|t|\le{n}}{\bf E}|TU^{0}_{G_{4}}(t)|
\end{equation*}
imply the following inequality
\begin{align}
\nonumber |\sum_{|t|\le{n}}M(x_{1},x_{2},x_{3},t)U(t,x_{4})|\le&{
\frac{v^{2}}{\eta^{2}}\sup_{|t|\le{n}}{\bf
E}|TU^{0}_{G_{4}}(t)|+\frac{v^{2}}{\eta^{2}}{\bf
E}|T|\sup_{|t|\le{n}}{\bf E}|U^{0}_{G_{4}}(t)|}\\
\nonumber &+\frac{2v^{2}}{\eta^{3}b}{\bf E}|T|
+\frac{6v^{2}}{\eta^{4}b^{2}}{\bf
E}|U^{0}_{G_{i}}(x_{i})U^{0}_{G_{j}}(x_{j})|\\
\label{b.5Wineq}
&+\frac{1}{\eta}\sup_{|t|\le{n}}|\Gamma_{1}(t)+\Gamma_{2}(t)|.
\end{align}
Henceforth, for sake of clarity, we consider
$G=G_{1}=G_{3}=\bar{G}_{2}=\bar{G}_{4}$ and $x=x_{r}$,
$r=1,\ldots,4$, then we get
$T=\left(U^{0}_{G}(x)\right)^{2}U^{0}_{\bar{G}}(x)$ and
\begin{equation}\label{b.5ET}
{\bf E}|T|\le{\left({\bf E}|U^{0}_{G}|^{4}\right)^{1/2}\left({\bf
E}|U^{0}_{G}|^{2}\right)^{1/2}}.
\end{equation}
Let us assume for the moment that
\begin{equation}\label{b.5supdaleth}
\sup_{|t|\le{n}}|\Gamma_{1}(t)+\Gamma_{2}(t)|=O\left(b^{-4}+
b^{-2}\sqrt{W}\right), \quad z\in\Lambda{\eta}
\end{equation}
with $W=\sup_{x}{\bf E}|U^{0}_{G}(x)|^{4}$. Now returning to
(\ref{b.5Wineq}) and gathering estimates given by relations
(\ref{b.4supEUG02}), (\ref{b.5ET}) and (\ref{b.5supdaleth}) imply
the following estimate
\begin{equation*}
W\le{A_{1}b^{-2}\sqrt{W}+A_{2}b^{-4}},
\end{equation*}
where $A_{1}$, $A_{2}$ are some constants. This proves
(\ref{b.4supEUG04}).

To complete the proof of Lemma 5.2, let us prove
(\ref{b.5supdaleth}). To do this, we use the following statement.
\vskip0,5cm

\begin{lem} If $z\in\Lambda_{\eta}$, then under conditions of
Theorem 2.1, the estimates
\begin{equation}\label{b.5DrU0Gx}
D^{r}_{st}\left(U^{0}_{G}(x)\right)=O(b^{-1}), \  \quad
r=1,\ldots,4,
\end{equation}
\begin{equation}\label{b.5DrT0Gts}
D^{r}_{st}\left(T^{0}\bar{G}(t,s)\right)=
O\left(b^{-3}+b^{-2}|U^{0}_{G}(x)|+b^{-1}|U^{0}_{G}(x)|^{2}
+|U^{0}_{G}(x)|^{3}\right), \quad r=3,4
\end{equation}
and
\begin{equation}\label{b.5EU0G2r}
{\bf E}|[U^{0}_{G}(x)]^{(\nu)}|^{2r}=O\left(b^{-3r}+{\bf
E}|U^{0}_{G}(x)|^{2r}\right), \ r=1,2
\end{equation}
hold for all $\nu=0,1,2$, all $|x|\le{n}$ and large enough $n$ and
$b$ satisfying (\ref{b.2bnalpha}).
\end{lem}

\vskip0,2cm We prove this Lemma at the end of this
subsection.\vskip0,2cm

Let us return to the proof of (\ref{b.5supdaleth}). Regarding the
variable $\Gamma_{1}$ (\ref{b.5daleth1}) and using
(\ref{b.4supEUG02}), (\ref{b.4K4estim}) and (\ref{b.5DrT0Gts}), one
gets with the help of (\ref{b.5ET}) that
\begin{equation}\label{b.5dalethestima1}
\sum_{|s|\le{n}}\frac{2[V_{4}+3v^{4}]}{3b}{\bf
E}|D^{3}_{st}(T^{0}\bar{G}(t,s))|U(s,t)=O\left(b^{-4}+
b^{-2}\sqrt{W}\right).
\end{equation}
Now let us estimate $\Gamma_{2}$ (\ref{b.5daleth2}). Regarding the
first term of the RHS of (\ref{b.5daleth2}) and using
(\ref{b.4supEUG02}), (\ref{b.5DrT0Gts}) and (\ref{b.5EU0G2r}), we
obtain inequality
\begin{align}
\nonumber &\sum_{|s|\le{n}}{\bf
E}|H(s,t)^{5}[D^{4}_{st}\left(T^{0}G_{4}(t,s)\right)]^{(0)}|\\
\nonumber &\le{c \sum_{|s|\le{n}}{\bf
E}\left\{\frac{|H(s,t)|^{5}}{b^{3}}+\frac{|H(s,t)|^{5}}{b^{2}}
|[U_{G}^{0}(x)]^{(0)}|+\frac{|H(s,t)|^{5}}{b}
|[U_{G}^{0}(x)]^{(0)}|^{2}\right\}}\\
\nonumber &+c\sum_{|s|\le{n}}{\bf
E}\left\{|H(s,t)|^{5}|[U_{G}^{0}(x)]^{(0)}|^{3} \right\}\\
\nonumber &\le{c
\sum_{|s|\le{n}}\left[\frac{\mu_{5}}{b^{11/2}}\psi\left(\frac{s-t}
{b}\right)+\frac{\mu_{10}^{1/2}}{b^{9/2}} \left({\bf
E}|[U_{G}^{0}(x)]^{(0)}|^{2}\right)^{1/2}\psi\left(\frac{s-t}
{b}\right)^{1/2}\right]}\\
\nonumber &+c \sum_{|s|\le{n}}\left[\frac{\mu_{10}^{1/2}}{b^{7/2}}
\left({\bf
E}|[U_{G}^{0}(x)]^{(0)}|^{4}\right)^{1/2}\psi\left(\frac{s-t}
{b}\right)^{1/2}\right] \\
\label{b.5daleth2estima1}
&=O\left(\frac{1}{b^{9/2}}+\frac{1}{b^{5/2}}\sqrt{W}\right).
\end{align}
Repeating the arguments used to prove (\ref{b.5daleth2estima1}), it
is easy to show that the term
\begin{equation*}
\sum_{|s|\le{n}}\frac{K_{2}}{3!}{\bf
E}\left\{H(s,t)^{3}[D^{4}_{st}\left(T^{0}G_{4}(t,s)\right)]^{(1)}
\right\}+\sum_{|s|\le{n}}\frac{K_{4}}{3!}{\bf
E}\left\{H(s,t)[D^{4}_{st}\left(T^{0}G_{4}(t,s)\right)]^{(2)}
\right\}
\end{equation*}
is of the order indicated in the RHS of (\ref{b.5supdaleth}) and
that
\begin{equation}\label{b.5dalethestima2}
\sup_{|t|\le{n}}|\Gamma_{2}(t)|=O\left(b^{-4}+
b^{-2}\sqrt{W}\right).
\end{equation}
Then the estimate (\ref{b.5supdaleth}) follows from
(\ref{b.5dalethestima1}) and (\ref{b.5dalethestima2}). Lemma 5.2 is
proved. \vskip0,2cm

{\it Proof of Lemma 5.3.} We prove Lemma 5.2 with $r=1$ because the
general case does not differ from this one. We start with the proof
of (\ref{b.5DrU0Gx}). Using (\ref{b.2partialG}), we obtain that
\begin{equation*}
D^{1}_{st}\left(U^{0}_{G}(x)\right)=-2\sum_{|k|\le{n}}G(k,s)G(k,t)
U(k,x).
\end{equation*}
Then estimate (\ref{b.5DrU0Gx}) (with $r=1$) follows from this
relation and inequality $|U(k,x)|\le{b^{-1}}$ and
(\ref{b.3sumG1ipG2ip}) (with $m=1$). The general case does not
differ from this one, so the estimate (\ref{b.5DrU0Gx}) is proved.

Let us prove (\ref{b.5DrT0Gts}). Remembering that
$T=[U^{0}_{G}(x)]^{2}U^{0}_{\bar{G}}(x)$ and using
(\ref{b.2partialG}) and (\ref{b.5DrU0Gx}), we obtain that
\begin{equation*}
D^{1}_{st}\{T^{0}\}=O(b^{-1}|U^{0}_{G}(x)|^{2}),
\end{equation*}
\begin{equation*}
D^{2}_{st}\{T^{0}\}=O\left(b^{-2}|U^{0}_{G}(x)|
+b^{-1}|U^{0}_{G}(x)|^{2}\right),
\end{equation*}
\begin{equation*}
D^{3}_{st}\{T^{0}\}=O\left(b^{-3}+b^{-2}|U^{0}_{G}(x)|
+b^{-1}|U^{0}_{G}(x)|^{2}\right).
\end{equation*}
Now it is easy to show that (\ref{b.5DrT0Gts}) is true.

\vskip0,2cm

Finally, we prove (\ref{b.5EU0G2r}) with $r=1$ because the general
case does not doffer from this one. To simplify computation, we use
the notation: for each pair $(s,t)$ and $\nu=0,1,2$, let
$H^{(\nu)}_{st}=H^{(\nu)}=\hat{H}$ be the matrix defined by
\begin{equation*}
\hat{H}(r,i)= \left\{
\begin{array}{lll}
H(r,i), & \textrm{if} & (r,i)\neq(s,t); \\
\hat{H}(s,t), &  \textrm{if} & (r,i)=(s,t)
\end{array}\right.
\end{equation*}
with $|\hat{H}(s,t)|\le{|H(s,t)|}$ and its resolvent by
$G^{(\nu)}_{sp}(z)=\hat{G}(z)$. Then the resolvent identity
(\ref{b.2h-zI-1}) imply that
\begin{align*}
U_{\hat{G}}(x)=&U_{G}(x)-\frac{1}{b}\sum_{|k|,|r|,|i|\le{n}}\hat{G}(k,r)
\{\hat{H}-H\}(r,i)G(i,k)\psi\left(\frac{x-k}{b}\right)\\
=&U_{G}(x)-\frac{1}{b}\sum_{|k|\le{n}}B(k,s,t)
\psi\left(\frac{x-k}{b}\right)
\end{align*}
with $B(k,s,t)=\hat{G}(k,s) [\hat{H}(s,t)-H(s,t)]G(t,k)$. Then
inequality (\ref{b.3GijleImz}) implies that
\begin{align*}
&{\bf E}|U^{0}_{\hat{G}}(x)|^{2}\\
&\le{2{\bf E}|U^{0}_{G}(x)|^{2}+\frac{2}{b^{2}} {\bf
E}\left|\sum_{|k|\le{n}}B^{0}(k,s,t)\psi\left(\frac{x-k}{b}\right)\right|^{2}}\\
&\le{2{\bf E}|U^{0}_{G}(x)|^{2}+\frac{8}{\eta^{4}b^{2}}{\bf
E}\left(|\hat{H}(s,t)|+{\bf E}|H(s,t)|\right)^{2}}\\
&\le{2{\bf E}|U^{0}_{G}(x)|^{2}+\frac{8}{\eta^{4}b^{3}}\left[{\bf
E}|a(s,p)|^{2}\psi\left(\frac{s-p}{b}\right)+3\left({\bf
E}|a(s,p)|\right)^{2}\psi\left(\frac{s-p}{b}\right)^{2}\right]}.
\end{align*}
This proves (\ref{b.5EU0G2r}). Lemma 5.3 is proved. $\hfill
\blacksquare$

\subsection{Proof of Lemma 4.3.} We prove relation
(\ref{b.4supbET12U}) with $k=1$ because the general case does not
differ from this one. To derive relations for the average value of
the variable $t_{12}(i,s)={\bf E}G_{1}(i,s)G_{2}(i,s)$, we use
identity (\ref{b.2h-zI-1}) and relation (\ref{b.2EXjFX1Xm}) (with
$q=3$) and repeat the proof of relation (\ref{b.4supbF12U}). Simple
computations lead to
\begin{align}
\nonumber t_{12}(i,s) = &\zeta_{2}g_{1}(i)\delta_{is} +
\zeta_{2}v^{2}t_{12}(i,s)U_{g_{2}}(s) \\
\label{b.5t12} &+
\zeta_{2}v^{2}\sum_{|p|\le{n}}t_{12}(i,p)g_{1}(s)U(p,s)+
\sum_{j=1}^{6}\gamma_{j}(i,s),
\end{align}
with
\begin{align*}
\gamma_{1}(i,s) =& \zeta_{2}v^{2}\sum_{|p|\le{n}}{\bf
E}\{G_{1}(i,s)G_{2}(i,p)G_{1}(p,s)\}U(p,s),\\
\gamma_{2}(i,s)=&\zeta_{2}v^{2}\sum_{|p|\le{n}}{\bf
E}\left\{G_{1}(i,s)G_{2}(i,p)G_{2}(p,s)\right\}U(p,s)\\
\gamma_{3}(i,s)=&\zeta_{2}v^{2}{\bf
E}\{G_{1}(i,s)G_{2}(i,s)U^{0}_{G_{1}}(s)\},\\
\gamma_{4}(i,s)=&\zeta_{2}v^{2}{\bf
E}\left\{G^{0}_{1}(s,s)\sum_{|p|\le{n}}G_{1}(i,p)G_{2}(i,p)U(p,s)
\right\},\\
\gamma_{5}(i,s)=&-\frac{\zeta_{2}}{6}\sum_{|p|\le{n}}K_{4}{\bf
E}\left\{D^{3}_{ps}(G_{1}(i,s)G_{2}(i,p))\right\}
\end{align*}
and
\begin{align*}
\gamma_{6}(i,s)=&-\frac{\zeta_{2}}{4!}\sum_{|p|\le{n}}{\bf
E}\left\{H(p,s)^{5}[D^{4}_{ps}\left(G_{1}(i,s)G_{2}(i,p)\right)]^{(0)}
\right\}\\
&+\frac{\zeta_{2}}{3!}\sum_{|p|\le{n}}K_{2}{\bf
E}\left\{H(p,s)^{3}[D^{4}_{ps}\left(G_{1}(i,s)G_{2}(i,p)\right)]^{(1)}
\right\}\\
&+\frac{\zeta_{2}}{3!}\sum_{|p|\le{n}}K_{4}{\bf
E}\left\{H(p,s)[D^{4}_{ps}\left(G_{1}(i,s)G_{2}(i,p)\right)]^{(2)}
\right\},
\end{align*}
where $K_{r}$, $r=2,4$ are the cumulants of $H(p,s)$ as in
(\ref{b.3K2K4}). Using (\ref{b.3sumG1ipG2ip}), it is easy to show
that
\begin{equation*}
\sup_{|i|,|s|\le{n}}|\gamma_{1}(i,s)|=o(b^{-1}), \quad
\sup_{|i|\le{n}}|\sum_{|s|\le{n}}\gamma_{1}(i,s)|=o(b^{-1}).
\end{equation*}
The same is valid for $\gamma_{2}$. Similar estimates for
$\gamma_{3}$, $\gamma_{4}$, $\gamma_{5}$ and $\gamma_{6}$ follow
from relations (\ref{b.4supEUG02}), (\ref{b.4supEGss2}) and simple
arguments as those to the proof of (\ref{b.4betaresti}) (see
(\ref{b.4K4estim})-(\ref{b.4beta5})). Thus, (\ref{b.5t12}) implies
that
\begin{equation}\label{b.5t12form1}
t_{12}(i,s) = g_{1}(i)q_{2}(i)\delta_{is} +
v^{2}g_{1}(s)q_{2}(s)\{t_{12}U\}(i,s) + \Delta(i,s),
\end{equation}
where
\begin{equation}\label{b.5Deltaestim}
\sup_{|i|,|s|\le{n}}|\Delta(i,s)|=o(1) \quad \hbox{ and } \quad
\sup_{|i|\le{n}}|\sum_{|s|\le{n}}\Delta(i,s)| = o(1)
\end{equation}
in the limit $n,b\rightarrow \infty$. We rewrite relation
(\ref{b.5t12form1}) in the matrix form (cf. (\ref{b.4F12form2}))
\begin{equation}\label{b.5t12form2}
t_{12} = \{ I - W^{(g_{1},q_{2})} \}^{-1}(Diag(g_{1}q_{2}) + \Delta
) = \sum_{m=0}^{+\infty}\{W^{(g_{1},q_{2})}\}^{m}(Diag(g_{1}q_{2}) +
\Delta ).
\end{equation}
Now we can apply to (\ref{b.5t12form2}) the same arguments as in the
proof of (\ref{b.4supbF12U}). Replacing $g_{1}$ and $q_{2}$ by
$w_{1}$ and $w_{2}$, respectively, we derive from
(\ref{b.5Deltaestim}) that for $i\in B_{L+Q}$,
\begin{equation}\label{b.5t12form3}
t_{12}(i,s) = \sum_{m=0}^{M}v^{2m}(w_{1}w_{2})^{m+1}[U^{m}](i,s) +
o(1),\quad n,b\rightarrow \infty.
\end{equation}
Multiplying both sides of (\ref{b.5t12form3}) by $U(s,i)$ and
summing over $s$, we obtain the relation
\begin{equation*}
\sum_{|s|\le{n}}t_{12}(i,s)U(s,i) =
\sum_{m=0}^{M}v^{2m}(w_{1}w_{2})^{m+1}[U^{m+1}](i,i) + o(1), \
N,b\rightarrow \infty.
\end{equation*}
Now convergence (\ref{b.4bUm+1}) implies the relation that leads,
with $M$ replaced by $\infty$, to (\ref{b.4supbET12U}). \vskip0,2cm

To prove (\ref{b.4supET12}), let us sum (\ref{b.5t12form3}) over
$s$. The second part of (\ref{b.5Deltaestim}) tells us that the
terms $\Delta$ remain small when summed over $s$. Thus we can write
relations
\begin{equation}\label{b.5sumt12is}
\sum_{|s|\le{n}}t_{12}(i,s)=
\sum_{m=0}^{M}(v^{2}w_{1}w_{2})^{m+1}\sum_{|s|\le{n}}[U^{m}](i,s) +
o(1) ,\ N,b\rightarrow \infty.
\end{equation}
Taking into account estimates for terms (\ref{b.4Pb})-(\ref{b.4Tnb})
(see previous work \cite{A} for more details), it is easy to observe
that convergence (\ref{b.4supUj-1}) together with
(\ref{b.5sumt12is}) imply (\ref{b.4supET12}). Finally, we prove
(\ref{b.4supEG2ii}). To derive relations for the average value of
variable $t_{11}(i,s)={\bf E}G_{1}(i,s)G_{1}(i,s)$, we repeat the
proof of (\ref{b.4supET12}) and replace $G_{2}$ by $G_{1}$. Then one
obtains (\ref{b.4supEG2ii}). Lemma 4.3 is proved. $\hfill
\blacksquare$ \vskip0,2cm

\section{Asymptotic properties of $T(z_{1},z_{2})$}

The asymptotic expression for $T(z_{1},z_{2})$ regarded in the limit
$z_{1}=\lambda_{1}+i0$, $z_{2}=\lambda_{2}+i0$ supplies one with the
information about the local properties of eigenvalue distribution
provided that $\lambda_{1}-\lambda_{2}=O(N^{-1})$. Indeed, according
to (\ref{b.2gnbz}), the formal definition of the eigenvalue density
$\rho_{n,b}(\lambda)=\sigma^{'}_{n,b}(\lambda)$ is
\begin{equation*}
\rho_{n,b}(\lambda)=\frac{1}{2i}[g_{n,b}(\lambda+i0)-g_{n,b}(\lambda-i0)].
\end{equation*}
We consider the density-density correlation function of $\rho_{n,b}$
\begin{equation*}
R_{n,b}(\lambda_{1},\lambda_{2})=-\frac{1}{4}\sum_{\delta_{1},\delta_{2}
=-1,1}\delta_{1}\delta_{2}C_{N,b}(\lambda_{1}+i\delta_{1}0,\lambda_{2}+
i\delta_{2}0).
\end{equation*}
In general, even if $R_{n,b}$ can be rigorously determined, it is
difficult to carry out direct study of it. Taking into account
relation $(2.13)$, one can simpler-expression
\begin{equation}\label{b.6Xi}
\Xi_{n,b}(\lambda_{1},\lambda_{2})=-\frac{1}{4Nb}
\sum_{\delta_{1},\delta_{2}=-1,+1}\delta_{1}\delta_{2}T(\lambda_{1}+
i\delta_{1}0,\lambda_{1}+i\delta_{1}0)
\end{equation}
and assume that it corresponds to the leading term to
$R_{n,b}(\lambda_{1},\lambda_{2})$ in the limit
$n,b\rightarrow\infty$.

It should be noted that for Wigner random matrices this approach is
justified by the study of the simultaneous limiting transition
$N\rightarrow\infty$, $\Im z_{j}\rightarrow0$ in the studies of
$C_{N}(z_{1},z_{2})$ \cite{DD,AAA,DDD,EE}.

\vskip0,5cm

\begin{theo} Let $T(z_{1},z_{2})$ is given by (\ref{b.2Tz1z2}). Assume
that function $\hat{\psi}(p)$ is such that there exist positive
constants $c_{1}$, $\delta$ and $v>1$ that
\begin{equation}\label{b.6hatpsicond}
\hat{\psi}(p)=\hat{\psi}(0)-c_{1}|p|^{\nu}+o(|p|^{\nu})
\end{equation}
for all $p$ such that $|p|\le{\delta}$, $\delta\rightarrow 0$. Then
\begin{equation}\label{b.6Xith}
\Xi_{n,b}(\lambda_{1},\lambda_{2})=\frac{1}{Nb}\frac{c_{2}}
{|\lambda_{1}-\lambda_{2}|^{2-1/v}}(1+o(1))
\end{equation}
for $\lambda_{j}$, $j=1,2$ satisfying
\begin{equation}\label{b.6lambdatheor}
\lambda_{1}, \lambda_{2}\rightarrow\lambda \in (-2v,2v).
\end{equation}
\end{theo}
\vskip0,2cm

We see from (\ref{b.2Tz1z2}) that there are two terms in
$T(z_{1},z_{2})$. The first was found in \cite{B} for band random
matrices, the second coincides with that found in \cite{C} for the
ensemble of Wigner random matrices. The proof of (\ref{b.6Xith})
consists of two parts already done in \cite{B} and \cite{C}. For
completeness, we reproduce here these computations.
 \vskip0,2cm

{\it Proof of Theorem 6.1}. Let us start with the term of
(\ref{b.6Xi}) that correspond to $\delta_{1}\delta_{2}=-1$. It
follows from (\ref{b.2wz}) that
\begin{equation}\label{b.61-v2w1w2}
\frac{1-v^{2}w_{1}w_{2}}{w_{1}w_{2}}=\frac{z_{1}-z_{2}}{w_{1}-w_{2}}.
\end{equation}
The above identity yields relations
\begin{equation*}
\epsilon|w(\lambda+i\epsilon)|^{2}=\Im w(\lambda+i\epsilon)(1-v^{2}|
w(\lambda+i\epsilon)|^{2}) \quad \hbox{ and } \quad
|w(\lambda+i0)|^{2}=v^{-2}
\end{equation*}
for $\lambda$ such that $\Im {w}(\lambda+i0)>0$. Combining these
relations with (\ref{b.1rho}) for the real and imaginary parts of
$w(\lambda+i0)=\tau(\lambda)+i\rho(\lambda)$, we obtain that
\begin{equation}\label{b.6taurho}
v^{2}\tau^{2}=\frac{\lambda^{2}}{4v^{2}} \quad \hbox{ and } \quad
v^{2}\rho^{2}=1-\frac{\lambda^{2}}{4v^{2}}
\end{equation}
(here and below we omit the variable $\lambda$). This implies the
existence of the limits $w(z_{1})=\overline{w(z_{2})}$ for
(\ref{b.6lambdatheor}). One can easily deduce from
(\ref{b.61-v2w1w2}) that in the limit (\ref{b.6lambdatheor})
\begin{equation}\label{b.61-v2w1w2estim}
\frac{1-v^{2}w(z_{1})w(z_{2})}{w(z_{1})w(z_{2})}=
\frac{\lambda_{1}-\lambda_{2}}{2i\rho}=o(1).
\end{equation}
Also we have that
\begin{equation}\label{b.6taurhorelation}
(1-v^{2}w^{2}_{1})(1-v^{2}w^{2}_{2})=2-2v^{2}(\tau^{2}-\rho^{2})=
4v^{2}\rho^{2}.
\end{equation}
Now let us consider the leading term of the correlation function.
Rewrite (\ref{b.2Tz1z2}) as
\begin{align*}
T(z_{1},z_{2})=&Q(z_{1},z_{2})+Q^{'}(z_{1},z_{2})
   +\frac{2v^{2}Q(z_{1},z_{2})}{(1-v^{2}w^{2}_{1})
   (1-v^{2}w^{2}_{2})}\\
=&\frac{2v^{2}S(z_{1},z_{2})}{(1-v^{2}w^{2}_{1}) (1-v^{2}w^{2}_{2})}
+Q^{'}(z_{1},z_{2})
\end{align*}
with
\begin{equation}\label{b.6S}
S(z_{1},z_{2})=\frac{1}{2\pi}\int_{-\infty}^{+\infty}
\frac{w^{2}_{1}w^{2}_{2}\hat{\psi}(p)}{( 1-
v^{2}w_{1}w_{2}\hat{\psi}(p))^{2}}dp
\end{equation}
and
\begin{equation}\label{b.6Q'}
Q^{'}(z_{1},z_{2})=\frac{2\Delta v^{4}w_{1}^{3}w_{2}^{3}}{(1 -
v^{2}w_{1}^{2})(1-v^{2}w_{2}^{2})},
\end{equation}
where $\Delta$ is given by (\ref{b.2Delta}). It is easy to observe
that relations (\ref{b.6taurho}) and
$|w(\lambda+i0)|^{2}=|w(\lambda-i0)|^{2}=1/v^{2}$ imply that (cf.
\cite{C})
\begin{equation}\label{b.6Q'proof}
Q^{'}(\lambda_{1}+i0,\lambda_{2}-i0)+Q^{'}(\lambda_{1}-i0,
\lambda_{2}+i0)=\frac{\Delta}{v^{4}\rho^{2}}.
\end{equation}
Now let us consider $S(z_{1},z_{2})$ (\ref{b.6S}) and let us write
\begin{equation*}
S(z_{1},z_{2})=\frac{1}{2\pi}\left\{\int_{-\delta}^{\delta}+
\int_{\mathbf{R}\setminus(-\delta,\delta)}\right\}
\frac{w^{2}_{1}w^{2}_{2}\hat{\psi}(p)} {(1-
v^{2}w_{1}w_{2}\hat{\psi}(p))^{2}}dp=I_{1}+I_{2}.
\end{equation*}
Relation (\ref{b.61-v2w1w2}) and (\ref{b.61-v2w1w2estim}) imply
equality
\begin{equation}\label{b.61-v2w1w2hatpsiestim}
[1- v^{2}w_{1}w_{2}\hat{\psi}(p)]^{2}=[\hat{\Psi}(p)-1]^{2}(1+o(1)).
\end{equation}
Since $\psi(t)$ is monotone, then
\begin{equation*}
\liminf_{p\in \mathbf{R}\setminus(-\delta,\delta)}[\Psi(p)-1]^{2}>0.
\end{equation*}
This means that $I_{2}<\infty$ in the limit (\ref{b.6lambdatheor}).
Relations (\ref{b.6hatpsicond}), (\ref{b.61-v2w1w2estim}) and
(\ref{b.61-v2w1w2hatpsiestim}) imply in the limit
(\ref{b.6lambdatheor}) and that if we take
$\delta|\lambda_{1}-\lambda_{2}|^{-1/\nu}\rightarrow\infty$, we
obtain asymptotically (cf. \cite{B})
\begin{equation}\label{b.6I1+I1}
I_{1}(\lambda_{1}+i0,\lambda_{2}-i0)+I_{1}(\lambda_{1}-i0,
\lambda_{2}+i0)=4B_{v}(c_{1})\frac{(2v\rho)^{2-1/\nu}}
{|\lambda_{1}-\lambda_{2}|^{2-1/\nu}},
\end{equation}
where
\begin{equation}\label{b.6Bvc1}
B_{v}(c_{1})=\frac{1}{2\pi
c^{1/\nu}_{1}}\left[\int_{0}^{\infty}\frac{ds}{1+s^{2\nu}}-2\int_{0}^
{\infty}\frac{ds}{(1+s^{2\nu})^{2}}\right]
\end{equation}
and $c_{1}$ is as in (\ref{b.6hatpsicond}). To prove
(\ref{b.6Xith}), it remains to consider the sum
\begin{equation*}
I_{1}(\lambda_{1}+i0,\lambda_{2}-i0)+I_{2}(\lambda_{1}-i0,
\lambda_{2}+i0).
\end{equation*}
It is easy to observe that relations of the form
(\ref{b.6taurhorelation}) imply the bounded ness of this sum in the
limit (\ref{b.6lambdatheor}).

Now gathering relations (\ref{b.6taurhorelation}),
(\ref{b.6Q'proof}) and (\ref{b.6I1+I1}), we derive that
\begin{equation}\label{b.6Xiproofresult}
\Xi_{n,b}(\lambda_{1},\lambda_{2})=\frac{1}{Nb}\frac{B_{\nu}(c_{1})}
{(2v\rho)^{1/\nu}}\frac{1}{|\lambda_{1}-\lambda_{2}|^{2-1/\nu}}(1+o(1)).
\end{equation}
This proves (\ref{b.6Xith}). $\hfill \blacksquare$ \vskip0,2cm

Let us discuss two consequences of Theorem 6.1.
\begin{itemize}
\item[$\bullet$] If $\nu=2$ and  $c_{1}=\int t^{2}\psi(t)<\infty$. Regarding the
RHS of (\ref{b.2Cnbz1z2Th}) in the limit (\ref{b.6lambdatheor}) with
$\lambda_{j}=\lambda+\frac{r_{j}}{N}$, $j=1,2$, we obtain the
asymptotic relation (see \cite{B})
\begin{align}
\nonumber
\Xi(\lambda_{1},\lambda_{2})&=-\frac{B_{2}(c_{1})}{2\sqrt{2}(v^{2}
\rho)^{1/2}}\frac{\sqrt{N}}{b}
\frac{1}{|r_{1}-r_{2}|^{3/2}}(1+o(1))\\
\label{b.6Xiform1}&=-C\frac{\sqrt{N}}{b}
\frac{1}{|r_{1}-r_{2}|^{3/2}}(1+o(1)), \quad C>0.
\end{align}

\item[$\bullet$] If $\Psi(t)=O(|t|^{-1-\nu})$ with $1<\nu<2$, we obtain the
asymptotic relation (see \cite{B})
\begin{equation}\label{b.6Xiform2}
\Xi(\lambda_{1},\lambda_{2})=\frac{B_{\nu}(c_{1})}{(2v^{2}\rho)^{1/\nu}}
\frac{N^{1-1/\nu}}{b}\frac{1}{|r_{1}-r_{2}|^{2-1/\nu}}(1+o(1))
\end{equation}
and conclude that the expression for (\ref{b.6Xi}) is proportional
to
\begin{equation*}
\frac{N^{1-1/\nu}}{b}\frac{1}{|r_{1}-r_{2}|^{2-1/\nu}}.
\end{equation*}

\end{itemize}

The form of asymptotic expressions (\ref{b.6Xiform1}) and
(\ref{b.6Xiform2}) coincides with the expressions determined by
Khorunzhy and Kirsch (see \cite{B}) for the spectral correlation
function of band random matrices \cite{B}.

The first conclusion is that the leading terms of the ensemble we
study (see (\ref{b.2Hnbij})) and the ensemble of band random
matrices are different but in the local scale, the form
(\ref{b.6Xiform1}) and (\ref{b.6Xiform2}) is the same. More
precisely, the tow ensembles mentioned above belong to the same
class of spectral universality.

Our main conclusion is that the limiting expression for $\Xi_{n,b}
(\lambda_{1},\lambda_{2})$ exhibits different behavior depending on
the rate of decay of $\psi(t)$ at infinity. In both cases (see
(\ref{b.6Xiform1}) and (\ref{b.6Xiform2})) the exponents do not
depend on the particular form of the function $\psi(t)$. Moreover,
in the first case the exponents do not depend on $\psi$ at all. This
can be regarded as a kind of spectral universality for the random
matrix ensembles $\{H_{n,b}\}$ (\ref{b.2Hnbij}). One can deduce that
these characteristics also do not depend on the probability
distribution of the random variables $a(i,j)$ (\ref{b.1ANij}).

\vskip0,5cm


\begin{thebibliography}{99}

\bibitem{A}  S. Ayadi: Semicircle Law For Random Matrices Of Long-Range
           Percolation Model. Arxiv PR/0806.4497v1, {\it to appear in Random Operators and Stochastic
           Eqs.} N4, Volume 16, (2009).


\bibitem{Aya2} S. Ayadi: Asymptotic properties of random matrices of long-range percolation
model. (submited in ROSE)


\bibitem{I}  D. Bessis, C. Itzykson, J. B. Zuber. Quantum field theory thechniques in graphical enumeration. {\it Adv. Appl. Math}. {\bf 1}, 109-157 (1980)


\bibitem{BleIts} P. Bleher and A. Its. Semiclassical asymptotics of
orthogonal polynomials, Rieman-Hilbert problem, and universality in
the matrix model. {\it Annals of Mathematics}, {\bf 150}, 185-266
(1999)



\bibitem{DD}  A. Boutet de Monvel, Khorunzhy: Asymptotic
             distribution of smoothed eigenvalue density:

             I. Gaussian random matrices, {\it Random Oper. Stoch. Eqs.}
             {\bf 7}, 1-22 (1999)

             II. Wigner random matrices, {\it Random Oper. Stoch. Eqs.}
             {\bf 7}, 149-167 (1999)




\bibitem{AAA} E. Brézin, A. Zee: Universality of the correlations
             between eigenvalues of large random matrices.{\it Nucl. Phys.} B
             {\bf 402} no. {\bf 3}, 613-627 (1993); Ambjorn J, Jurkiewicz J,
             Makeenko Yu M.:Multiloop correlators for two-dimensional quantum
             gravity. {\it Phys. lett.} B {\bf 251} (4), 517-524 (1990)



\bibitem{Casati} G. Casati, L. Molinari, F. Izrailev. Scaling
properties of band random matrices. {\it Phys. Rev. Lett.} {\bf 64}
1851 (1990)



\bibitem{K}    D. Coppersmith, D. Gamarnik, M. I. Sviridenko: The diametre of
              long-range percolation graph. {\it In Mathematics and Computer
              Science II}. {\it Trends Math.,
              Birkhauser, Basel}, 147-159 (2002)

\bibitem{II}  A. Crisanti, G. Paladin, A. Vulpiani. Products of Random Matrices in Statistical Physics. {\it Berlin: Springer}, (1993)



\bibitem{DDD} F. J. Dyson: Statistical theory of the energy levels of
             complex systems (III).{\it J.Math. Phys} {\bf 3}, 166-175 (1962)



\bibitem{Deift} P. A. Deift, A. Its, X. Zhou. A Riemann-Hilbert approach
to asymptotic problems arising in the theory of random matrix
models, and also in the theory of integrable statistical mechanics.
{\it Ann. Math} {\bf 146}, 149-235 (1997)


\bibitem{III}  T. Guhr, A. Müller-Groeling, H. A. Weidenmüller. Random -matrix                    theories in quantum physics: {\it Common concepts, Phys. Rep.}                  {\bf 299}, 189-425 (1998)


\bibitem{Fyodorv} Y. V. Fyodorov, A. D. Mirlin. Scaling properties of
localization in random band matrices. A $\sigma$ model approach.
{\it Phys. Rev. Lett.} {\bf 67}, 2405 (1991)



\bibitem{IIII}  F. Haake. Quantum Signatures of Chaos. {\it Berlin: Springer},                  (1991)


\bibitem{BBBB}  S. Janson, T. Luczak, A. Rucinski. Random Graphs.  John Wiles and                 Sons, {\it Inc. New York}. (2002)

\bibitem{B}  A. Khorunzhy, W. Kirsch: On Asymptotic Expansions and
           Scales of Spectral Universality in Band Random Matrix
           Ensembles.{\it Commun. Math. Phys.} {\bf 231}, 223-255 (2002)




\bibitem{EEEE}  O. Khorunzhiy, W. Kirsch, P. Müller. Lifshitz tails
for spectra of Erd\H os-R\`enyi random graphs. {\it Ann. Appl.
Probab}. Volume 16, Number 1, 295-309, (2006)


\bibitem{EE}  A. Khorunzhy: On smoothed density of states for Wigner
             random matrices. {\it Rand. Oper. Stoch. Eqs.} {\bf 5}, 147-162
             (1997)


\bibitem{C}  A. Khorunzhy, B. Khoruzhenko, L. Pastur: Asymptotic
           properties of large random matrices with independent
           entries.{\it J.Math. Phys}. {\bf 37} , 5033-5060 (1996)


\bibitem{CCC}  A. Khorunzhy, B. Khoruzhenko, L. Pastur, M. Shcherbina.
              Large-n limit in statistical mechanics and the spectral theory
              of disordered systems. {\it In Phase Transitions and Critical
              Phenomena}, Vol.15, edg C.Domb and J.L.Lebowitz.Academic Press,
              London, pp. 73-239, (1992)



\bibitem{D} A. Khorunzhy, L. Pastur: On the eigenvalue distribution of
           the deformed Wigner ensemble of random matrices. {\it Adv.
           Soviet. Math}. {\bf 19}, 97-107 (1994)




\bibitem{FFF}  V. Marchenko, L. Pastur. {\it Math. USSR-sb} {\bf 1},
             457 (1967)

\bibitem{FF}  V. Marchenko, L. Pastur: Eigenvalue distribution of
             some class of random matrices. {\it Matem. Sbornik}. {\bf 72},
             507 (1972)



\bibitem{CC}  M. L. Mehta: Random matrices, 2nd ed. {\it Academic, New
             York}, (1991)

\bibitem{AA}  S. A. Molchanov, L. Pastur, A. Khorunzhy: Eigenvalue
             distribution for band random matrices in the limit of
             their infinite rank. {\it Teor. Matem. Fizika} {\bf 99},
             (1992)

\bibitem{EEE} L. A. Pastur. {\it Theor. Math.Phys.} {\bf 10}, 67 (1972)


\bibitem{BBB} C. Porter: Statistical Theories of Spectra: Fluctuations.
             {\it New York: Acad. Press}, (1965)



\bibitem{Soshni} A. B. Soshnikov. Universality at the edge of the
spectrum in Wigner random matrices. {\it Commun. Math. Phys}. {\bf
207}, 697-733 (1999)




\bibitem{syl} P. Sylvestrov. Summing graphs for random band matrices.
{\it Phys. Rev.} E {\bf 55}, 6419-6432 (1997)



\bibitem{Voi} D. Voiculescu, K. J. Dykema, A. Nica. Free Random
Variables, A noncommutative probability approch to free products
with applications to random matrices, operator algebras and harmonic
analysis on free groups. CRM Monograph Series, {\bf 1}. Providence,
RI: AMS, 1992




\bibitem{BB}  E. Wigner: Charecteristic vector of bordered matrices
             with infinite dimentions. {\it Ann. Math.} {\bf 62}, (1955)




\end{thebibliography}
\end{document}